\documentclass[12pt]{article}
\usepackage{amsmath}
\usepackage{epsfig}
\usepackage{amssymb}
\usepackage{hyperref}
\usepackage{bbold}
\usepackage{braket}
\usepackage{dsfont}

\usepackage{mathtools}
\usepackage{amsfonts}
\usepackage[numbers,sort&compress]{natbib}

\def\be{\begin{equation}}
\def\ee{\end{equation}}
\def\beq{\begin{equation}}
\def\eeq{\end{equation}}
\def\beqa{\begin{eqnarray}}
\def\eeqa{\end{eqnarray}}
\def\ba{\begin{eqnarray}}
\def\ea{\end{eqnarray}}
\def\bea{\begin{eqnarray}}
\def\eea{\end{eqnarray}}

\newcommand\as{\alpha_s}
\newcommand\f[2]{\frac{#1}{#2}}

\def\beq{\begin{equation}}
\def\eeq{\end{equation}}
\def\beeq{\begin{eqnarray}}
\def\eeeq{\end{eqnarray}}
\def\to{\rightarrow}
\def\nn{\nonumber}

\def\b0{b_0}

\def\b0{b_0}

\begin{document}

\begin{titlepage}
\renewcommand{\thefootnote}{\fnsymbol{footnote}}
\begin{flushright}
     \end{flushright}
\par \vspace{10mm}
\begin{center}
{\large \bf
Approximate NNLO QCD corrections to semi-inclusive DIS}\\

\vspace{8mm}

\today
\end{center}

\par \vspace{2mm}
\begin{center}
{\bf Maurizio Abele${}^{\,a}$,}
\hskip .2cm
{\bf Daniel de Florian${}^{\,b}$,}
\hskip .2cm
{\bf Werner Vogelsang${}^{\,a}$  }\\[5mm]
\vspace{5mm}
${}^{a}\,$ Institute for Theoretical Physics, T\"ubingen University, 
Auf der Morgenstelle 14, \\ 72076 T\"ubingen, Germany\\[2mm]
${}^b$ International Center for Advanced Studies (ICAS) and ICIFI, ECyT-UNSAM,\\
Campus Miguelete, 25 de Mayo y Francia, (1650) Buenos Aires, Argentina \\
\end{center}


\vspace{9mm}
\begin{center} {\large \bf Abstract} \end{center}
We determine approximate next-to-next-to-leading order (NNLO) corrections to unpolarized and 
polarized semi-inclusive DIS. They
are derived using the threshold resummation formalism, which we fully develop to next-to-next-to-leading
logarithmic (NNLL) accuracy, including the two-loop hard factor. The approximate NNLO terms are obtained by expansion 
of the resummed expression. They include all terms in Mellin space that are logarithmically enhanced at threshold, 
or that are constant. In terms of the customary SIDIS variables $x$ and $z$ they include all double distributions 
(that is, ``plus'' distributions and $\delta$-functions) in the partonic variables. We also investigate 
corrections that are suppressed at threshold and we determine the dominant terms among these. 
Our numerical estimates suggest much significance of the approximate NNLO terms, along with a
reduction in scale dependence.

\end{titlepage}  

\setcounter{footnote}{2}
\renewcommand{\thefootnote}{\fnsymbol{footnote}}

\section{Introduction \label{intro}}

Data taken in the semi-inclusive deep-inelastic scattering (SIDIS) process $\ell p\to \ell hX$
offer powerful insights into QCD and hadronic structure. Among their main uses are extractions
of fragmentation functions~\cite{deFlorian:2017lwf,Anderle:2016czy,Leader:2015hna,Bertone:2018ecm,Khalek:2021gxf}, 
(polarized) parton distributions~\cite{deFlorian:2009vb,DeFlorian:2019xxt}, or even combinations thereof~\cite{Ethier:2017zbq,Moffat:2021dji}. 

Today, modern ``global'' analyses of parton distributions are customarily carried out at next-to-next-to-leading order (NNLO)
accuracy of QCD perturbation theory. Although SIDIS might in principle offer important complementary information on,
for example, the flavor structure of the sea quarks, the analyses usually do not include information from SIDIS.
One reason for this is the fact that the NNLO partonic hard-scattering functions for SIDIS are not yet available
(a few first steps toward their calculation have been taken in~\cite{Daleo:2003xg,Daleo:2003jf,Anderle:2016kwa}),
so that computations of the SIDIS cross section are currently restricted to next-to-leading order (NLO). 

The Electron Ion Collider (EIC) is now firmly on its path toward construction~\cite{AbdulKhalek:2021gbh}.
The past few years have seen tremendous progress on the development of the theoretical framework for describing
reactions relevant at the EIC. Further improvements will likely occur in the near term. Ideally, by the time the EIC will turn on,
it would be hoped that the precision of theoretical calculations should be on par with 
what has by now been achieved for the LHC, with NNLO corrections available for many
observables, and extractions of parton distributions and fragmentation functions routinely at NNLO using numerically 
efficient tools. As part of this, it is expected that also full calculations of the NNLO corrections to SIDIS will become 
available at some point. Until this is the case, it is useful to provide accurate approximations of the NNLO corrections 
for SIDIS. This is the main goal of this paper. The results we obtain may be used to carry out analyses of 
parton distributions and/or fragmentation functions using SIDIS data at (approximate) NNLO already now.

The strategy we will follow to derive approximate NNLO corrections to SIDIS is to use QCD threshold resummation. 
The partonic SIDIS process is characterized by two ``scaling'' variables, $\hat{x}=-q^{2}/2 p \cdot q\equiv Q^{2}/2 p \cdot q$ 
and $\hat{z}=p\cdot p_c/p \cdot q$,
with $q,p,p_c$ the momenta of the virtual photon, the incoming parton, and the fragmenting parton, respectively. 
When $\hat{x}$ and $\hat{z}$ get close to 1, the partonic hard-scattering functions develop large double-logarithmic
terms. These logarithms arise since large $\hat{x},\hat{z}$ corresponds to scattering near a phase space boundary, 
where real-gluon emission is suppressed. At the $k$th order of perturbation theory, the SIDIS quark hard-scattering function
contains terms of the form 
$\alpha_s^k\delta(1-\hat{x}) \left(\frac{\ln^{m}(1-\hat{z})}{1-\hat{z}}\right)_+$,
$\alpha_s^k\delta(1-\hat{z}) \left(\frac{\ln^{m}(1-\hat{x})}{1-\hat{x}}\right)_+$, with
$m\leq 2k-1$,
or ``mixed'' terms $\alpha_s^k \left(\frac{\ln^{m}(1-\hat{x})}{1-\hat{x}}\right)_+
\left(\frac{\ln^{n}(1-\hat{z})}{1-\hat{z}}\right)_+$ with $m+n\leq 2k-2$. Here the subscript ``+'' indicates
the usual distribution. Threshold resummation addresses these large logarithmic terms to all orders
in the strong coupling. The resummation for the case of SIDIS was 
discussed in Refs.~\cite{Cacciari:2001cw,Anderle:2012rq,Anderle:2013lka,Sterman:2006hu}
to next-to-leading logarithm (NLL), which amounts to the cases $m=2k-1,2k-2,2k-3$ and 
$m+n= 2k-2,2k-3$ above, respectively. To NLO, this reproduces {\it all} double distributions, 
but only the three leading towers of logarithms at NNLO and beyond.

In the present paper we take a significant step further and extend the work of~\cite{Anderle:2012rq,Anderle:2013lka}
to  next-to-next-to-leading logarithm (NNLL). The close correspondence of SIDIS with the
Drell-Yan cross section is particularly useful in this context~\cite{Sterman:2006hu}, and so is
the close correspondence between the totally inclusive Drell-Yan cross section and the
cross section differential in rapidity~\cite{Anderle:2012rq,Westmark:2017uig,Banerjee:2018vvb}. 
We use results available in the literature~\cite{Catani:2013tia,Catani:2014uta,Gehrmann:2005pd,Moch:2005tm} 
to determine the two-loop hard virtual contribution to the resummed expression for SIDIS.
The NNLL results may then be expanded to fixed order, NNLO. 
The main important new result of our paper is that we derive {\it all} double distributions in $\hat{x}$ and $\hat{z}$ in the NNLO
SIDIS quark coefficient function. 
We further improve our results by deriving the dominant part of NNLO contributions that are
suppressed near threshold. These terms are of the form $\ln^m(1-\hat{x}) \ln^n(1-\hat{z})$, with $m+n=3$. 
We also show that the NNLO contributions near threshold are the same for the spin-averaged
and spin-dependent cases. This is indeed expected for the terms with $+\,$-distributions, since  
these terms are associated with emission of soft gluons which does not care about spin,
but it extends even to the threshold-suppressed contributions that we derive. 

Our results are readily suited for phenomenology for the SIDIS cross section and spin asymmetry
at (nearly) NNLO.  At the very least, they provide important benchmarks for future full calculations
of SIDIS at NNLO. 

Our paper is organized as follows: Section~\ref{pSIDIS} sets the stage by addressing
the perturbative SIDIS cross section. In Sec.~\ref{thresh} we determine the threshold resummation for
SIDIS to NNLL. Special emphasis is put on the derivation of the hard factor at two loops.
Section~\ref{sec:NLPbyevol} addresses the dominant threshold-suppressed contributions.
Having determined all ingredients, we finally present the NNLO expansions in Sec.~\ref{NNLOexp}. 
Section~\ref{Pheno} rounds off the paper by presenting some basic phenomenological results
at approximate NNLO.

\section{Perturbative SIDIS cross section \label{pSIDIS}}

We consider the semi-inclusive deep-inelastic scattering (SIDIS) process
$\ell(k)\, p(P)\rightarrow \ell'(k')\, h(P_{h}) \, X$ with the momentum transfer $q = k-k'$.
It is described by the variables
\beeq\label{kindef}
Q^{2} 
	&= &-q^{2} = - (k-k')^{2}\; , \nn\\[2mm]
x 
	&= &\frac{Q^{2}}{2 P \cdot q} \; ,\nn\\[2mm]
y
	&= &\frac{P\cdot q}{P \cdot k}\; , \nn\\[2mm]
z
	&= &\frac{P \cdot P_{h}}{P \cdot q}\; .
\eeeq
We have $Q^{2} = xys$, with $\sqrt{s}$ the center-of-mass (c.m.) energy for the incoming electron and proton.
We may write the spin-averaged SIDIS cross section as (see, for example~\cite{Anderle:2012rq})
\beq\label{sidiscrsec}
\frac{d^{3} \sigma^{h}}{dxdydz}\,=\,\frac{4 \pi \alpha^{2}}{Q^2} \left[ 
		\frac{1+(1-y)^{2}}{2y} \mathcal{F}^{h}_{T}(x,z,Q^2)
		+ \frac{1-y}{y} \mathcal{F}^{h}_{L}(x,z,Q^2)\right]\,,
\eeq
where $\alpha$ is the fine structure constant and ${\cal F}_T^h\equiv 2F_1^h$ 
and ${\cal F}_L^h\equiv F_L^h/x$ are the transverse and longitudinal structure 
functions. For collisions of longitudinally polarized leptons and protons we obtain
the helicity-dependent cross section as 
\beeq
\frac{1}{2}\left( \frac{d^{3} \sigma_{++}^h}{dxdydz} - \frac{d^{3} \sigma_{+-}^h}{dxdydz} \right)
	&\equiv&
\frac{d^{3} \Delta \sigma^h}{dxdydz} \,=\,
	\frac{4 \pi \alpha^{2}}{Q^2} \,\frac{1-(1-y)^{2}}{2y}\, \mathcal{G}^{h}_1(x,z,Q^2)\,,
\eeeq
with $ \mathcal{G}^{h}_1=2 g_1^h$ in the more conventional notation of Ref.~\cite{deFlorian:1997zj}.
The subscripts in the first expression denote the helicities of the incoming lepton and proton.

Using factorization, the unpolarized structure functions may be written as
\begin{equation}
\label{F1hallorders}
\mathcal{F}^{h}_{i}(x,z,Q^2)
	\,=\, \sum_{f,f'} \int_{x}^{1} \frac{d\hat{x}}{\hat{x}} \int_{z}^{1} \frac{d \hat{z}}{\hat{z}}\,
	D^{h}_{f'}\left( \frac{z}{\hat{z}},\mu_{F}\right)
	\omega^{i}_{f'f} \left(\hat{x},\hat{z},\alpha_{s}(\mu_{R}), \frac{\mu_{R}}{Q}, \frac{\mu_{F}}{Q} \right)
	f\left( \frac{x}{\hat{x}},\mu_{F}\right)\,,
\end{equation}
for $i=T,L$. Here $f(\xi,\mu_F)$ is the distribution of parton $f=q,\bar{q},g$ in the nucleon
at momentum fraction $\xi$ and factorization scale $\mu_F$, while $D^h_{f'} \left(\zeta,\mu_F\right)$
is the corresponding fragmentation function for parton $f'$ going to the
observed hadron\footnote{We always use the same factorization scales in the initial and the final state.} $h$.
The functions $\omega^i_{f'f}$ are the spin-averaged hard-scattering
coefficient functions. In the same way, we have in the spin-dependent case:
\beq
\label{g1hallorders}
\!\!\mathcal{G}_1^h(x,z,Q^2) \,=\,\sum_{f,f'} 
\int_x^1 \frac{d\hat{x}}{\hat{x}}\int_z^1 \frac{d\hat{z}}{\hat{z}}\, 
D^h_{f'} \left(\frac{z}{\hat{z}},\mu_F\right)\,\Delta \omega_{f'f}
\left(\hat{x},\hat{z},\alpha_{s}(\mu_{R}), \frac{\mu_{R}}{Q}, \frac{\mu_{F}}{Q}\right) \,
\Delta f \left(\frac{x}{\hat{x}},
\mu_F\right) \,,\;\;\;\;
\eeq
with the proton's spin-dependent parton distribution functions $\Delta f$, and 
with spin-dependent hard-scattering functions $\Delta \omega_{f'f}$. 

The $\omega^i_{f'f},\Delta \omega_{f'f}$ can be computed in QCD perturbation theory. 
Their expansions read
\beq\label{Cpert}
\omega^i_{f'f}\,=\,\omega^{i,(0)}_{f'f}+\frac{\alpha_s(\mu_R)}{\pi}\,\omega^{i,(1)}_{f'f}
+\left(\frac{\alpha_s(\mu_R)}{\pi}\right)^2 \omega^{i,(2)}_{f'f}+
{\cal O}(\alpha_s^3)\,,
\eeq
and
\beq\label{DCpert}
\Delta \omega_{f'f}\,=\,\Delta \omega^{(0)}_{f'f}+\frac{\alpha_s(\mu_R)}{\pi}\Delta \omega^{(1)}_{f'f}
+\left(\frac{\alpha_s(\mu_R)}{\pi}\right)^2\Delta \omega^{(2)}_{f'f}+
{\cal O}(\alpha_s^3)\,.
\eeq
Here the strong coupling is evaluated at the renormalization scale $\mu_R$.
To lowest order (LO), only the process $\gamma^*q\to q$ contributes, and we have 
\ba\label{LO}
\omega^{T,(0)}_{qq}(\hat{x},\hat{z})\,=\,\Delta \omega^{(0)}_{qq}(\hat{x},\hat{z})&=& e_q^2\,
\delta(1-\hat{x})\delta(1-\hat{z}),\nn\\[2mm]
\omega^{L,(0)}_{qq}(\hat{x},\hat{z})&=&0,
\ea
with the quark's fractional charge $e_q$. Beyond LO, also gluons in the
initial or final state contribute. The  first-order coefficient functions $(\Delta)\omega^{i,(1)}_{f'f}$
have been known for a long time~\cite{altarelli,Nason:1993xx,Furmanski:1981cw,graudenz,deFlorian:1997zj,deFlorian:2012wk}
(see also Ref.~\cite{Anderle:2012rq}).

In the following, it is convenient to take Mellin moments of the SIDIS cross section, 
for which the convolutions in Eqs.~(\ref{F1hallorders}),(\ref{g1hallorders}) turn
into ordinary products. We define
\beq\label{moms}
\tilde{{\cal F}}^h_i(N,M,Q^2)\equiv\int_0^1 dx\, x^{N-1}\int_0^1 dz\, z^{M-1}\,
{\cal F}^h_i(x,z,Q^2)\,,
\eeq
and in the same way for $\mathcal{G}_1^h$. One readily finds from~(\ref{F1hallorders})
\beq\label{Melmo}
\tilde{{\cal F}}^h_i(N,M,Q^2)\,=\,\sum_{f,f'}\tilde{D}_{f'}^{h}(M,\mu_F)\,
\tilde{\omega}^i_{f'f}\left(N,M,\alpha_{s}(\mu_{R}), \frac{\mu_{R}}{Q}, \frac{\mu_{F}}{Q}\right)\tilde{f}(N,\mu_F)\,,
\eeq
where
\ba
&&\tilde{f}(N,\mu_F)\equiv\int_0^1 dx \,x^{N-1}f(x,\mu_F),\nn\\[2mm]
&&\tilde{D}_{f'}^{h}(M,\mu_F)\equiv\int_0^1 dz \,z^{M-1}D^h_{f'}(z,\mu_F),\nn\\[2mm]
&&\tilde{\omega}^i_{f'f}\left(N,M,\alpha_{s}(\mu_{R}), \frac{\mu_{R}}{Q}, \frac{\mu_{F}}{Q}\right)\,\equiv\,
\int_0^1 d\hat{x}\,\hat{x}^{N-1}\int_0^1 d\hat{z}\,\hat{z}^{M-1}\,
\omega^i_{f'f}\left(\hat{x},\hat{z},\alpha_{s}(\mu_{R}), \frac{\mu_{R}}{Q}, \frac{\mu_{F}}{Q}\right) \,.\hspace*{6mm}
\ea
We observe that the Mellin moments of the structure functions are obtained from 
the moments of the parton distribution functions and fragmentation
functions, and the double-Mellin moments of  the partonic hard-scattering functions.
For the spin-dependent case we have in the same way
\beq\label{Melmo1}
\tilde{{\cal G}}^h_1(N,M,Q^2)\,=\,\sum_{f,f'}\tilde{D}_{f'}^{h}(M,\mu_F)\,
\Delta \tilde{\omega}_{f'f}\left(N,M,\alpha_{s}(\mu_{R}), \frac{\mu_{R}}{Q}, \frac{\mu_{F}}{Q}\right)\,
\Delta \tilde{f}(N,\mu_F)\,,
\eeq
with the corresponding moments $\Delta \tilde{f}(N,\mu_F)$ and $\Delta \tilde{\omega}_{f'f}(N,M,\ldots)$ 
of the polarized parton distributions and hard-scattering functions, respectively. 

For the perturbative expansions given in Eqs.~(\ref{Cpert}),(\ref{DCpert}), we have at lowest order according to~(\ref{LO})
\ba
\tilde{\omega}^{T,(0)}_{qq}(N,M)\,=\,\Delta \tilde{\omega}^{(0)}_{qq}(N,M)&=&e_q^2 ,\nn\\[2mm]
\tilde{\omega}^{L,(0)}_{qq}(N,M)&=&0.
\ea
The corresponding moments of the next-to-leading order (NLO) terms $\omega^{i,(1)}_{f'f},\Delta \omega^{(1)}_{f'f}$
may be found in Refs.~\cite{Stratmann:2001pb,Anderle:2012rq}. In the following, we address higher-order
corrections to the hard-scattering functions that arise at large values of $\hat{x}$ and $\hat{z}$
or, equivalently, at large $N$ and $M$. 

\section{Threshold resummation \label{thresh}}

\subsection{Structure of resummation for Drell-Yan and SIDIS}

As has been discussed in~\cite{Anderle:2012rq,Anderle:2013lka} (and as is familiar from numerous 
other situations in perturbative calculations of cross sections), in the ``threshold limit'' of large $N$ and $M$ the perturbative 
QCD corrections for $\tilde{\omega}^T_{f'f}$ and $\Delta \tilde{\omega}_{f'f}$ develop large double-logarithmic corrections in $\ln(N)$ and
$\ln(M)$. These corrections exponentiate and may thus be controlled to all orders in the strong coupling, amounting
to a resummation of the logarithmic corrections. The exponentiated result may be used to obtain approximate fixed-order 
corrections to the SIDIS cross sections.

To achieve the resummation of the threshold logarithms for SIDIS, we will use the methods developed in 
Refs.~\cite{Catani:2003zt,Sterman:2006hu,Hinderer:2018nkb,Anderle:2012rq}. Technically, the resummation for 
SIDIS with its two Mellin variables $N$ and $M$ bears much resemblance with that for the Drell-Yan or Higgs cross sections
at measured rapidity, which are also described by two separate moments~\cite{Catani:1989ne,Cacciari:2001cw,Westmark:2017uig,Banerjee:2018vvb}.
This is in contrast to observables characterized by a {\it single} moment variable $N$, such as the totally inclusive Drell-Yan
cross section. However, as was shown in Ref.~\cite{Anderle:2012rq} to NLL, there is a simple correspondence
between the threshold-resummed expressions for the case with two Mellin moments, and those with only a single moment.
To state this correspondence, let us consider the resummed $q\bar{q}$ hard-scattering function for the Drell-Yan process as an example.
For the totally inclusive cross section, we denote the function by $\tilde{\omega}^{\mathrm{DY,incl}}_{q\bar{q}}(N)$, where $N$ is 
the Mellin variable conjugate to $z\equiv Q^2/\hat{s}$, with $Q$ the Drell-Yan pair mass and $\sqrt{\hat{s}}$ the partonic c.m. energy. 
For the rapidity-dependent cross section, we have instead $\tilde{\omega}^{\mathrm{DY,rap}}_{q\bar{q}}(N,M)$, where
$N$ and $M$ are conjugate to $\sqrt{z}\,{\mathrm{e}}^{\pm y}$, respectively, with $y$ the lepton pair's rapidity. Near threshold one
then has
\beq
\tilde{\omega}^{\mathrm{DY,rap}}_{q\bar{q}}(N,M)\,=\,\tilde{\omega}^{\mathrm{DY,incl}}_{q\bar{q}}(\sqrt{NM})\,.
\eeq
This correspondence is a consequence of kinematics in the exponentiation of eikonal diagrams
as discussed in Refs.~\cite{Anderle:2012rq,Sterman:2006hu,Laenen:2000ij}. It applies to all
color-singlet processes and may therefore also be exploited for the SIDIS process\footnote{As discussed in Ref.~\cite{Sterman:2006hu},
one may actually define a simplified variant of SIDIS that is characterized by only a single Mellin variable, conjugate to
$\tau_{\mathrm{SIDIS}}=xz$, with $x,z$ defined in Eq.~(\ref{kindef}).}. As a result, we may obtain resummed expressions 
for SIDIS by considering those for the inclusive Drell-Yan process, ``rescaling'' $N$ to $\sqrt{NM}$ appropriately,
and ``crossing'' from timelike (Drell-Yan) kinematics to spacelike (SIDIS) kinematics. This is the strategy we will pursue in this paper. 

There are various (of course, equivalent) ways of writing the all-order expression for the resummed inclusive Drell-Yan 
hard-scattering function near threshold. Here we will follow the approaches developed in Refs.~\cite{Catani:2003zt,Hinderer:2018nkb}.
We have, in the $\overline{\mathrm{MS}}$ scheme,
\beeq\label{DYres}
&&\hspace*{-1cm}\tilde{\omega}^{\mathrm{DY,res}}_{q\bar{q}}\left( N,\alpha_{s}(\mu_{R}), \frac{\mu_{R}}{Q}, \frac{\mu_{F}}{Q}\right)\,=\,
e_q^2  \,H^{\mathrm{DY}}_{q\bar{q}} \left(\alpha_{s}\big(\mu_{R}\big),\frac{\mu_{R}}{Q}, \frac{\mu_{F}}{Q}
\right)\,\Delta_{q}\left(N,\alpha_{s}(\mu_{R}), \frac{\mu_{R}}{Q}, \frac{\mu_{F}}{Q}\right)\nn\\[2mm]
&=&e_q^2  \,H^{\mathrm{DY}}_{q\bar{q}} \left(\alpha_{s}\big(\mu_{R}\big),\frac{\mu_{R}}{Q}, \frac{\mu_{F}}{Q}
\right) \,	\widehat{C}_{qq} \left(\alpha_{s}\big(\mu_{R}\big),\frac{\mu_{R}}{Q}
\right)\nn\\[2mm]
&\times& \exp\left\{\int_{Q^{2}/\bar{N}^2}^{Q^2} 
	\frac{d \mu^{2}}{\mu^{2}} \left[  A_{q}\big(\alpha_{s}(\mu)\big) \ln\left( \frac{\mu^{2} \bar{N}^2}{Q^{2}}\right)
		- \frac{1}{2} \widehat{D}_{q} \big(\alpha_{s} (\mu)\big)\right]
		\,+\, 2 \ln\bar{N} \int_{Q^2}^{ \mu^{2}_{F}} \frac{d \mu^{2}}{\mu^{2}} A_{q}\big(\alpha_{s}(\mu)\big)\right\},\nn\\
\eeeq
where
\beq\label{Nbar}
\bar{N}\,=\,N\,{\mathrm{e}}^{\gamma_E}\,,
\eeq
with the Euler constant $\gamma_E$. 
In Eq.~(\ref{DYres}) each of the functions $H^{\mathrm{DY}}_{q\bar{q}},\widehat{C}_{qq},A_{q}, \widehat{D}_{q}$ is a perturbative series in the
strong coupling with expansion coefficients that are collected in Appendix~\ref{sec:appendixAndim} to the order required for
resummation at next-to-next-to-leading logarithmic (NNLL)  accuracy. The factor $\Delta_{q}$ in the first line contains 
all soft-gluon radiation near threshold (both collinear and wide-angle), while the coefficient $H^{\mathrm{DY}}_{q\bar{q}}$
collects hard virtual corrections to the underlying lowest-order (LO) process (here, $q \bar{q}\to \gamma^{*}$), which are 
independent of the moment variable. In the second line we have followed Refs.~\cite{Catani:2003zt,Hinderer:2018nkb}
to split up the soft-gluon factor $\Delta_{q}$ into the term $\widehat{C}_{qq}$ that is again independent of $N$,
and an exponential that contains all $N$-dependence. The latter is in fact entirely a function of $\ln(\bar{N})$ and
contains no further $N$-independent terms. 

The NNLL resummation formula for the SIDIS transverse structure function may now be written as follows:
\beq\label{SIDISres}
\tilde{\omega}^{T,{\mathrm{res}}}_{qq}\left( N,M,\alpha_{s}(\mu_{R}), \frac{\mu_{R}}{Q}, \frac{\mu_{F}}{Q}\right)\,=\,
e_q^2  \,H^{\mathrm{SIDIS}}_{qq} \left(\alpha_{s}\big(\mu_{R}\big),\frac{\mu_{R}}{Q}, \frac{\mu_{F}}{Q}
\right)\,\Delta_{q}\left(\sqrt{NM},\alpha_{s}(\mu_{R}), \frac{\mu_{R}}{Q}, \frac{\mu_{F}}{Q}\right)\,.
\eeq
As  anticipated, we have ``rescaled'' $N$ to $\sqrt{NM}$ in the moment-dependent part of the expression. The 
function $\Delta_{q}$ is otherwise identical to that for the Drell-Yan case in Eq.~(\ref{DYres}), including
the function $\widehat{C}_{qq}$. The hard
coefficient $H^{\mathrm{SIDIS}}_{qq}$ is, however, different from $H^{\mathrm{DY}}_{q\bar{q}}$, owing
to the different kinematics of the two processes. It will be derived in the next subsection. Inserting $\Delta_{q}$ from~(\ref{DYres}) into
Eq.~(\ref{SIDISres}) we obtain
\beeq\label{SIDISres1}
\tilde{\omega}^{T,{\mathrm{res}}}_{qq}\left( N,M,\alpha_{s}(\mu_{R}), \frac{\mu_{R}}{Q}, \frac{\mu_{F}}{Q}\right)&=&
e_q^2  \,H^{\mathrm{SIDIS}}_{qq}  \left(\alpha_{s}\big(\mu_{R}\big),\frac{\mu_{R}}{Q}, \frac{\mu_{F}}{Q}
\right) \,	\widehat{C}_{qq} \left(\alpha_{s}\big(\mu_{R}\big),\frac{\mu_{R}}{Q}\right) \nn\\[2mm]
&\times&\exp\left\{\int_{Q^{2}/(\bar{N}\bar{M})}^{Q^2} 
	\frac{d \mu^{2}}{\mu^{2}} \left[  A_{q}\big(\alpha_{s}(\mu)\big) \ln\left( \frac{\mu^{2} \bar{N}\bar{M}}{Q^{2}}\right)
		- \frac{1}{2} \widehat{D}_{q} \big(\alpha_{s} (\mu)\big)\right]\right.\nn\\[2mm]
	&+&\left. \ln\bar{N} \int_{Q^2}^{ \mu^{2}_{F}} \frac{d \mu^{2}}{\mu^{2}} A_{q}\big(\alpha_{s}(\mu)\big)
	\,+\,\ln\bar{M} \int_{Q^2}^{ \mu^{2}_{F}} \frac{d \mu^{2}}{\mu^{2}} A_{q}\big(\alpha_{s}(\mu)\big)\right\}\,,
\eeeq
where (see~(\ref{Nbar})) $\bar{M}=M{\mathrm{e}}^{\gamma_E}$. We note that the same resummation formula
applies to the spin-dependent case:
\beq
\Delta \tilde{\omega}^{\mathrm{res}}_{qq}\,=\,\tilde{\omega}^{T,{\mathrm{res}}}_{qq}\,.
\eeq

\subsection{The hard factor $H^{\mathrm{SIDIS}}_{qq}$}

As already mentioned, the factor $H^{\mathrm{SIDIS}}_{qq}$ is derived from the finite part of the virtual corrections to the LO process, which 
for SIDIS is $q\gamma^*\to q$. Since we want to derive the resummed formula to NNLL (and ultimately the near-threshold 
NNLO corrections to SIDIS), we need $H^{\mathrm{SIDIS}}_{qq}$ to two loops. The relevant two-loop virtual corrections are known
in terms of the ``quark form factor'' computed to two and even three loops in Refs.~\cite{Gehrmann:2005pd,Moch:2005tm,Gehrmann:2010ue}.
In case of the space-like kinematics ($q^2<0$) relevant for SIDIS the renormalized spacelike quark form factor is given to two loops 
in dimensional regularization with $d=4-2\epsilon$ space-time dimensions as~\cite{Gehrmann:2005pd,Moch:2005tm}
\begin{equation}
	F_{q}(q^{2})
		\,=\, F^{(0)}_{q} +  \frac{\alpha_{s}}{\pi} \, F^{(1)}_{q}
			+ \left( \frac{\alpha_{s}}{\pi}  \right)^{2} F^{(2)}_{q}
			+ \mathcal{O}(\alpha_{s}^{3})\,,
\end{equation}
where 
\beeq
F^{(0)}_{q} &= &1 \,,\nn \\[2mm]
F^{(1)}_{q}
	&=&C_{F}\left[-\frac{1}{2 \epsilon^2} 
		-\frac{3}{4 \epsilon }
		+\frac{\pi ^2}{24}-2
		+\left(\frac{7 \zeta (3)}{6}+\frac{\pi ^2}{16}-4\right) \epsilon  \right.\nn\\[2mm]
	&+&\left. \left(\frac{7 \zeta (3)}{4}+\frac{47 \pi ^4}{2880}+\frac{\pi ^2}{6}-8\right) \epsilon^2
	+\mathcal{O}\left(\epsilon ^3\right)\right]\,,\nn\\[2mm]
F^{(2)}_{q}&=&C^{2}_{F}
\left[\frac{1}{8 \epsilon ^4}+\frac{3}{8 \epsilon ^3}+\left(\frac{41}{32}-\frac{\pi ^2}{48}\right)\frac{1}{\epsilon^2}+\left(\frac{221}{64}-\frac{4 \zeta (3)}{3}\right)\frac{1}{\epsilon} 
\right.\nn\\[2mm]
&-&\left. \frac{29 \zeta (3)}{8}-\frac{13 \pi ^4}{576}+\frac{17 \pi ^2}{192}+\frac{1151}{128}\right]\nn \\[2mm]
&+& C_{F} C_{A} \left[ \frac{11}{32 \epsilon ^3}+\left( \frac{1}{9}+\frac{\pi ^2}{96} \right)\frac{1}{\epsilon^2}+\left(\frac{13 \zeta (3)}{16}-\frac{11 \pi ^2}{192}-\frac{961}{1728}\right)\frac{1}{\epsilon }  \right.\nn \\[2mm]
&+&\left.\frac{313 \zeta (3)}{144}+\frac{11 \pi ^4}{720}-\frac{337 \pi ^2}{1728}-\frac{51157}{10368}\right] \nn\\[2mm]
&+& C_{F} N_{f}  \left[-\frac{1}{16 \epsilon ^3}-\frac{1}{36 \epsilon^2}+\left( \frac{65}{864}+\frac{\pi ^2}{96} \right)\frac{1}{\epsilon }+
\frac{\zeta (3)}{72}+\frac{23 \pi ^2}{864}+\frac{4085}{5184} \right]+\mathcal{O}\left(\epsilon\right)\,,
\eeeq
with $N_f$ the number of flavors and $C_F=4/3,C_A=3$. 
In these expressions we have kept terms of order $\epsilon$ and $\epsilon^2$ in the one-loop result since these turn
out to make finite contributions in the end. 

As shown in Refs.~\cite{Catani:2013tia,Catani:2014uta}, the hard coefficient may be extracted from the form factor in the following way. 
Applied to the case of SIDIS we have from~\cite{Catani:2014uta}
\begin{equation}
H_{qq}^{\mathrm{SIDIS}} \big(\alpha_{s}(Q)\big)\,=\, 
\left|\, 	\big[ 1 - \tilde{I}_q\big(\epsilon, \alpha_s(Q)\big) \big] F_q\,\right|^2	\,, 
\label{eq:Cthdef}
\end{equation}
where $\tilde{I}_q$ is an operator that removes the poles of the form factor and makes the necessary
soft and collinear adjustments needed to extract the hard coefficient. It is given in~\cite{Catani:2014uta} in 
terms of a convenient all-order form:
\begin{equation}
1 - \tilde{I}_q (\epsilon, \alpha_s)\,=\, 
\exp\left\{ R_q\left(\epsilon, \alpha_{s}\right) - i \Phi_q\left(\epsilon,\alpha_{s}\right)\right\}\,,
\end{equation}
with functions $R_q$ and $\Phi_q$ that each are perturbative series. The phase $\Phi_q$ does not contribute
in our case since we take the absolute square in Eq.~(\ref{eq:Cthdef}). The function $R_q$ effects the
cancelation of infrared divergences from the quark form factor. It can be expressed in terms 
of a soft and a collinear part:
\begin{equation}
R_q(\epsilon, \alpha_{s})
	=  R^{\mathrm{\,soft}}_q(\epsilon, \alpha_{s}) + R^{\mathrm{\,coll}}_q(\epsilon, \alpha_{s})\,,
\end{equation}
where for NNLL accuracy
\beeq
R_q^{\mathrm{\,soft}}(\epsilon, \alpha_{s}) 
	&=& C_F \left( \frac{\alpha_{s}}{\pi}R_q^{\mathrm{\,soft}\, (1)}(\epsilon) + \left(\frac{\alpha_{s}}{\pi}\right)^{2} 
	R_q^{\mathrm{\,soft}\, (2)}(\epsilon) + \mathcal{O}(\alpha^{3}_{s})  \right)\; ,\nn\\[2mm]
R^{\mathrm{\,coll}}_q(\epsilon, \alpha_{s})
	&=&\frac{\alpha_{s}}{\pi}R_q^{\mathrm{\,coll}\, (1)}(\epsilon) + \left(\frac{\alpha_{s}}{\pi}\right)^{2} 
	R_q^{\mathrm{\,coll}\, (2)}(\epsilon) + \mathcal{O}(\alpha^{3}_{s})\,,
\eeeq
with
\beeq
R_q^{\mathrm{\,soft}\, (1)}(\epsilon)
	&=& \frac{1}{2 \epsilon^{2}} -\frac{\pi ^2}{8} \; ,\nn \\[2mm]
R_q^{\mathrm{\,soft}\, (2)}(\epsilon)
	&=& - \frac{3\pi b_0}{8\epsilon^{3}} + \frac{1}{8 \epsilon^{2}} \,\frac{A^{(2)}_{q}}{C_{F}} \nn\\[2mm]
	&-& \frac{1}{16 \epsilon} \left[ C_{A} \left(7 \zeta(3)+\frac{11 \pi ^2}{36}-\frac{202}{27}\right)+ N_{f}\left(\frac{28}{27}-\frac{\pi ^2}{18}\right)\right] \nn\\[2mm]
	&+&C_{A} \left(-\frac{187 \zeta(3)}{144}+\frac{\pi ^4}{288}-\frac{469 \pi ^2}{1728}+\frac{607}{648}\right)+ N_{f} \left(\frac{17 \zeta(3)}{72}+\frac{35 \pi ^2}{864}-\frac{41}{324}\right)	\; , \nn \\[2mm]	
R_q^{\mathrm{\,coll}\, (1)}(\epsilon)
	&= &\frac{3}{4 \epsilon} C_{F} \; ,\nn\\[2mm]
R_q^{\mathrm{\,coll}\, (2)}(\epsilon)
	&=& -\frac{3 \pi b_0}{8\epsilon^{2}}C_{F}+ \frac{1}{8\epsilon} \left[ C_{F}^2 \left(6 \zeta(3)-\frac{\pi ^2}{2}+\frac{3}{8}\right)
	+\, C_{A} C_{F} \left(-3 \zeta(3) +\frac{11 \pi ^2}{18}+\frac{17}{24}\right)  \right. \nonumber\\[2mm]
	&&\left. \hspace*{2.9cm}+\,C_{F} N_{f}\left(-\frac{1}{12}-\frac{\pi ^2}{9}\right)\right] .
\eeeq
The coefficient $b_{0}$ can be found in Appendix~\ref{sec:appendixAndim}. Inserting all terms into Eq.~(\ref{eq:Cthdef}) and expanding in $\alpha_s$, all poles in powers of $1/\epsilon$ cancel, and we 
find for an arbitrary renormalization scale $\mu_R$, but for $\mu_F=Q$:
\begin{equation}
H_{qq}^{\mathrm{SIDIS}} \left(\alpha_{s}(\mu_R), \frac{\mu_{R}}{Q}, 1\right)\,=\,	 1 + \frac{\alpha_{s}(\mu_R)}{\pi} \,H_{qq}^{\mathrm{SIDIS},(1)} 
			+ \left( \frac{\alpha_{s}(\mu_R)}{\pi} \right)^{2} H_{qq}^{\mathrm{SIDIS},(2)} 
			+ \mathcal{O}(\alpha^{3}_{s})\,,
\end{equation}
with
\beeq\label{H12sidis}
H_{qq}^{\mathrm{SIDIS},(1)} 
	&=& C_{F}\left(-4-\frac{\pi ^2}{6}\right)\,,\nn \\[2mm]
H_{qq}^{\mathrm{SIDIS},(2)} 
	&= & C_{F}\left(- 4 - \frac{\pi ^2}{6}\right)\pi b_{0}\ln \frac{\mu^{2}_{R}}{Q^{2}}
		+C^{2}_{F}\left(-\frac{15  \zeta (3)}{4}+\frac{61 \pi ^2 }{48}+\frac{511}{64}-\frac{\pi ^4 }{60}\right)\nn\\[2mm]
	&+&C_{F}C_{A}\left(\frac{7 \zeta (3)}{4}+\frac{3 \pi ^4 }{80}-\frac{1535 }{192}-\frac{403 \pi ^2 }{432} \right)
		+C_{F} N_{f}\left(\frac{ \zeta (3)}{2}+\frac{29 \pi ^2}{216}+\frac{127 }{96} \right)\,.
\eeeq
The factorization scale dependence of $H_{qq}^{\mathrm{SIDIS}}$ is trivially determined by the DGLAP
evolution kernels of the parton distributions and fragmentation functions and will be addressed later.

With all ingredients to NNLL resummation at hand we are now also in the position to expand the
hard-scattering function in~(\ref{SIDISres1}) to NNLO (that is, ${\cal O}(\alpha_s^2)$) accuracy. 
This expansion will be carried out in Sec.~\ref{NNLOexp}. Before turning to it, we
will discuss another class of corrections near threshold that are suppressed with respect to the terms
addressed by resummation, but that can be significant as well in phenomenological studies.

\section{Subleading contributions near threshold}
\label{sec:NLPbyevol}

All contributions contained in Eq.~(\ref{SIDISres1}) are {\it leading} near threshold in the
sense that they carry powers of $\ln(N)$ or $\ln(M)$, never accompanied by any suppression by 
$1/N$ or $1/M$. Such terms are therefore often referred to as {\it leading-power (LP)} contributions. 
For the NNLL resummed cross section the LP terms contain the five ``towers'' $\alpha_s^n L^m$, with 
$m\in\{2n,\ldots,2n-4\}$, where $L^m$ can be any product of (in total) $m$ logarithms in $N$ or $M$.
The LP terms correspond to distributions (``$+$''-distributions and $\delta$-functions) in $\hat{x},\hat{z}$ space. 
In the full cross section there are, of course, also terms that are suppressed near threshold.
The most important among these are terms still containing logarithms, but suppressed by a single
power in $1/N$ or $1/M$. Such terms are known as {\it next-to-leading power (NLP)} corrections. 
Their structure is $\alpha_s^n L^m/N$ or $\alpha_s^n L^m/M$, with $m\in\{2n-1,\ldots,2n-3\}$, corresponding to terms
of the form $\alpha_s^n \ell^m$ in $\hat{x},\hat{z}$ space, 
where $\ell^m$ is a product of $\ln(1-\hat{x})$ and $\ln(1-\hat{z})$ with total power $m$. 

The role of NLP terms in color-singlet hard-scattering cross sections has
been addressed early on in Refs.~\cite{Kramer:1996iq,Akhoury:1998tb,Kulesza:2002rh,Catani:2003zt,Shimizu:2005fp}. 
In recent years, the understanding of such corrections has further advanced,
and numerous studies have been carried out~\cite{Laenen:2008ux,Grunberg:2009yi,Laenen:2010uz,Bonocore:2014wua,Bonocore:2015esa,Moch:2009mu,Vogt:2010cv,Almasy:2010wn,LoPresti:2014ihe,Almasy:2015dyv,DelDuca:2017twk,Moult:2018jjd,Ebert:2018gsn,Bahjat-Abbas:2019fqa,vanBeekveld:2019cks,vanBeekveld:2019prq,Ajjath:2020ulr,Ajjath:2020sjk,Ajjath:2020lwb,Ajjath:2021lvg,Beneke:2019oqx,Broggio:2021fnr,Liu:2020tzd,Lustermans:2019cau} that address the NLP contributions from various angles, such as corrections to the eikonal approximation, resummations of
NLP terms to leading logarithm and beyond, and generalized factorization theorems at NLP. For especially simple processes such 
as the fully inclusive Drell-Yan process, the results of these studies are quite mature. For processes described by two scaling variables (or, two
Mellin moments), as relevant for SIDIS, comparably fewer studies are available~\cite{Lustermans:2019cau,Ajjath:2020lwb}. In the present study we will
derive the dominant NLP contributions at NNLL which, as described above, are of the form $\alpha_s^n L^{2n-1}/N$ or $\alpha_s^n L^{2n-1}/M$. 
In terms of the NNLO expansion, these are the terms $\alpha_s^2 L^3/N$ or $\alpha_s^2 L^3/M$, where 
$L\in\{\ln^3(N),\ln^2(N)\ln(M),\ln(N)\ln^2(M),\ln^3(M)\}$. 

As discussed in~\cite{Kramer:1996iq,Kulesza:2002rh,Catani:2003zt}, these dominant NLP terms may be incorporated to
all orders via a particular treatment of the evolution of the parton distributions and fragmentation functions\footnote{For
an alternative, but equivalent, approach in $\hat{x},\hat{z}$ space, see Ref.~\cite{vanBeekveld:2019cks}.}. To this end,
we consider a specific SIDIS quark channel in the spin-averaged case and include the parton distribution and fragmentation function.
From Eqs.~(\ref{Melmo}),(\ref{SIDISres1}) the corresponding resummed contribution to the transverse SIDIS structure function in moment space
may be written as
\beeq\label{SIDISres3}
&&\hspace*{-1.4cm}\tilde{q}(N,\mu_F)\,\tilde{D}_q^{h}(M,\mu_F)\,
\tilde{\omega}^{T,{\mathrm{res}}}_{qq}\left( N,M,\alpha_{s}(\mu_{R}), \frac{\mu_{R}}{Q}, \frac{\mu_{F}}{Q}\right)\nn\\[2mm]
&&=\;e_q^2  \,H^{\mathrm{SIDIS}}_{qq} \left(\alpha_{s}(\mu_R), \frac{\mu_{R}}{Q},\frac{\mu_{F}}{Q} \right)\,
\exp\left\{ -2\int^{Q^2}_{ \mu^{2}_{F}} \frac{d \mu^{2}}{\mu^{2}} P_{q,\delta} \big(\alpha_{s}(\mu)\big)
\right\}\,\widehat{C}_{qq}
\left(\alpha_{s}(\mu_R), \frac{\mu_{R}}{Q} \right)\nn\\[2mm]
&&\times\;	\exp\left\{\int_{Q^{2}/(\bar{N}\bar{M})}^{Q^2} 
	\frac{d \mu^{2}}{\mu^{2}} \left[  A_{q}\big(\alpha_{s}(\mu)\big) \ln\left( \frac{\mu^{2}}{Q^{2}}\right)+
	2P_{q,\delta} \big(\alpha_{s}(\mu)\big)
		- \frac{1}{2} \widehat{D}_{q} \big(\alpha_{s} (\mu)\big)\right]\right\}\nn\\[2mm]
&&\times\;	\exp\left\{ \int^{Q^2/(\bar{N}\bar{M})}_{ \mu^{2}_{F}} \frac{d \mu^{2}}{\mu^{2}} \Big[ -A_{q}\big(\alpha_{s}(\mu)\big)\,\ln\bar{N} + P_{q,\delta} \big(\alpha_{s}(\mu)\big)\Big]
\right\}\,\tilde{q}(N,\mu_F)	\nn\\[2mm]
&&\times\;	\exp\left\{ \int^{Q^2/(\bar{N}\bar{M})}_{ \mu^{2}_{F}} \frac{d \mu^{2}}{\mu^{2}} \Big[ -A_{q}\big(\alpha_{s}(\mu)\big)\,\ln\bar{M} + P_{q,\delta} \big(\alpha_{s}(\mu)\big)\Big]
\right\}\,\tilde{D}_q^{h}(M,\mu_F)\,,
\eeeq
where the function $P_{q,\delta}$ corresponds to the coefficient of $\delta(1-x)$ in the quark DGLAP splitting function and is also given in 
Appendix~\ref{sec:appendixAndim}.  

We make the following observations concerning Eq.~(\ref{SIDISres3}). We obviously have simply added and subtracted the terms involving $P_{q,\delta}$
in the exponent, so that they cancel. However, each of the individual terms serves a separate purpose. The $P_{q,\delta}$ term in the second line,
when combined with $H^{\mathrm{SIDIS}}_{qq} \big(\alpha_{s}(\mu_R), \mu_R/Q, \mu_F/Q\big)$, removes the factorization scale dependence of 
the SIDIS hard function, so that we end up with $H^{\mathrm{SIDIS}}_{qq} \big(\alpha_{s}(\mu_R), \mu_R/Q,1\big)$, precisely as given in Eq.~(\ref{H12sidis}). 
Thanks to factorization, this must hold true to all orders of perturbation theory. The other two $P_{q,\delta}$ terms in Eq.~(\ref{SIDISres3}) 
combine with the terms $A_{q}\ln\bar{N}$ or $A_{q}\ln\bar{M}$ to reproduce the quark-to-quark splitting function in the large-$N$ or large-$M$ limit, 
at leading power. As a result, the last two exponential factors simply represent the DGLAP evolutions of the quark parton 
distribution function and the fragmentation function, respectively, from scale $\mu_F$ to scale $Q/\sqrt{\bar{N}\bar{M}}$. At leading power,
this evolution is entirely diagonal, and evolution of parton distributions (spacelike) and of fragmentation functions (timelike) is identical.
We can therefore carry out this evolution and write Eq.~(\ref{SIDISres3}) as
\beeq\label{SIDISres4}
&&\hspace*{-1.2cm}\tilde{q}(N,\mu_F)\,\tilde{D}_q^{h}(M,\mu_F)\,
\tilde{\omega}^{T,{\mathrm{res}}}_{qq}\left( N,M,\alpha_{s}(\mu_{R}), \frac{\mu_{R}}{Q}, \frac{\mu_{F}}{Q}\right)\nn\\[2mm]
&&\hspace*{-0.5cm}=\;e_q^2  \,H^{\mathrm{SIDIS}}_{qq} \left(\alpha_{s}(\mu_R), \frac{\mu_{R}}{Q},1\right)\,\widehat{C}_{qq}
\left(\alpha_{s}(\mu_R), \frac{\mu_{R}}{Q} \right)\,\tilde{q}\left(N,Q/\sqrt{\bar{N}\bar{M}}\right)	\,\tilde{D}_q^{h}\left(M,Q/\sqrt{\bar{N}\bar{M}}\right)\,,
\nn\\[2mm]
&&\hspace*{-0.5cm}\times\;	\exp\left\{\int_{Q^{2}/(\bar{N}\bar{M})}^{Q^2} 
	\frac{d \mu^{2}}{\mu^{2}} \left[  A_{q}\big(\alpha_{s}(\mu)\big) \ln\left( \frac{\mu^{2}}{Q^{2}}\right)+
	2P_{q,\delta} \big(\alpha_{s}(\mu)\big)
		- \frac{1}{2} \widehat{D}_{q} \big(\alpha_{s} (\mu)\big)\right]\right\}\,.
\eeeq
Again, this is correct to all orders. The trick now to obtain the dominant NLP corrections is to evolve the parton 
distributions and fragmentation functions from scale $Q/\sqrt{\bar{N}\bar{M}}$ back to scale $\mu_F$, but now 
using the DGLAP evolution {\it including} NLP corrections~\cite{Kramer:1996iq,Kulesza:2002rh,Catani:2003zt}. 
The latter are readily obtained from the $1/N$ or $1/M$ terms in the spacelike or timelike splitting functions, respectively. 
As it turns out, for the {\it dominant} NLP terms, only the $1/N$ (or $1/M$) terms in the {\it leading-order}
splitting kernels need to be taken into account. The related terms in the higher-order splitting functions 
lead to contributions that have fewer logarithms. Let us for the moment continue to consider only diagonal evolution,
corresponding to the SIDIS quark channel. 
We write the standard LO quark-to-quark splitting function at large values of the moment variable as
\beeq
\mathcal{P}_{qq}^N&=&\frac{\alpha_s}{\pi}\left( -A_{q}^{(1)}\ln\bar{N} + P_{q,\delta}^{(1)} + \frac{Q_q^{(1)}}{N}\right)+{\cal O}(\alpha_s^2)
\nn\\[2mm]
&=&\frac{\alpha_s}{\pi}\,C_F \left(-\ln\bar{N} +\frac{3}{4} -\frac{1}{2N}\right)+{\cal O}(\alpha_s^2)\,.
\eeeq
The term proportional to $Q_q^{(1)}$ is the NLP correction. At this order, the spacelike and timelike quark-to-quark splitting 
functions are identical so that also their NLP corrections are the same. The relation between 
$\tilde{q}\left(N,Q/\sqrt{\bar{N}\bar{M}}\right)$ and $\tilde{q}(N,\mu_F)$ including the dominant NLP correction is now given by
\beq\label{evolnew}
\tilde{q}\left(N,\frac{Q}{\sqrt{\bar{N}\bar{M}}}\right)=\exp\left\{ \int^{Q^2/(\bar{N}\bar{M})}_{ \mu^{2}_{F}} \frac{d \mu^{2}}{\mu^{2}}
\left[ -A_{q}\big(\alpha_{s}(\mu)\big)\ln\bar{N} + P_{q,\delta} \big(\alpha_{s}(\mu)\big)
+\frac{\alpha_{s}(\mu)}{\pi}\, \frac{Q_q^{(1)}}{N}\right]\right\}\tilde{q}(N,\mu_F),
\eeq
and in the same way for the quark fragmentation functions. As we discussed, only the LO term with $Q_q^{(1)}$ is
relevant for the dominant NLP corrections. The terms with $A_{q}$ and $P_{q,\delta}$ remain, of course, the full
all-order functions, needed to second order (NLO) for our purpose of obtaining NNLL/NNLO accuracy. 
We see from Eq.~(\ref{evolnew}) that the dominant NLP corrections in the quark channel are obtained by
multiplying the full resummed expression in Eq.~(\ref{SIDISres3}) by the two factors
\beq\label{nlpfactors}
\exp\left\{-\int^{Q^2/(\bar{N}\bar{M})}_{ \mu^{2}_{F}} \frac{d \mu^{2}}{\mu^{2}}
\frac{\alpha_{s}(\mu)}{\pi}\, \frac{C_F}{2N}\right\}\,
\exp\left\{-\int^{Q^2/(\bar{N}\bar{M})}_{ \mu^{2}_{F}} \frac{d \mu^{2}}{\mu^{2}}
\frac{\alpha_{s}(\mu)}{\pi}\, \frac{C_F}{2M}\right\}\,,
\eeq
corresponding to the NLP terms related to diagonal evolution of the parton distribution and the fragmentation function.

As is well known, once the NLP terms are included, the evolution of parton distributions and fragmentation functions
also involves quark-gluon mixing and hence is no longer diagonal, taking instead a 
matrix form. Transitions among quarks of different flavor turn out to be suppressed as $1/N^2$ or higher,
at least through NLO in the evolution kernels which is all we need here. Including the dominant NLP corrections,
the full evolution equations for the parton distributions may be cast into the form
\beq\label{cast}
\frac{d}{d\ln\mu^2} \, \left(\begin{matrix}\tilde{q}\big(N, \mu \big) \\[1mm] \tilde{g}\big(N, \mu\big) \end{matrix}\right)\,=\,
\mathcal{P}_s^N\big(\alpha_{s}(\mu) \big)
\left(\begin{matrix}\tilde{q}\big(N, \mu\big) \\[1mm] \tilde{g}\big(N, \mu\big) \end{matrix}\right)
 \eeq
to all orders, where $\mathcal{P}_s^N(\alpha_{s})$ denotes the NLO matrix of spacelike splitting functions in moment space,
which may be found in~\cite{Floratos:1981hs}. A corresponding equation holds for the fragmentation functions, 
with however the timelike splitting functions $\mathcal{P}_t^M$~\cite{Floratos:1981hs}. 

It is interesting to explore the implications of the singlet mixing and to see what NLP effects it generates beyond
the quark-to-quark channel. We will do this as part of the NNLO expansion to be discussed in the next section.
For this expansion we do not need to fully solve
the evolution equation (although this could be done using the techniques of Ref.~\cite{Furmanski:1981cw}). 
Instead, it suffices to just solve the equation to second order in the strong coupling, which may be achieved by iterating the kernel:
\beeq\label{evolit}
&&\hspace*{-0.8cm}\left(\begin{matrix}\tilde{q}\left(N,Q/\sqrt{\bar{N}\bar{M}}\right) \\[2mm] \tilde{g}\left(N,Q/\sqrt{\bar{N}\bar{M}}\right)\end{matrix}\right)\nn\\[2mm]
&&=\; \left( \mathds{1} + \int_{ \mu^{2}_{F}}^{\frac{Q^{2}}{\bar{N}\bar{M}}} \frac{d q^{2}}{q^{2}} \;  \mathcal{P}_s^N\big(\alpha_{s}(q)\big)
	+ \int_{ \mu^{2}_{F}}^{\frac{Q^{2}}{\bar{N}\bar{M}}} \frac{d q^{2}}{q^{2}} \;  \mathcal{P}_s^N\big(\alpha_{s}(q)\big)
	\int_{ \mu^{2}_{F}}^{\frac{q^{2}}{\bar{N}\bar{M}}} \frac{d \tilde{q}^{2}}{\tilde{q}^{2}} \;  \mathcal{P}_s^N\big(\alpha_{s}(\tilde{q})\big)
	\right)\left(\begin{matrix}\tilde{q}\big(N, \mu_F\big) \\[2mm] \tilde{g}\big(N,  \mu_F\big)\end{matrix}\right)\,,\nn\\
	\label{eq:evolmatrix}
\eeeq
and similarly for the fragmentation functions. This expression may then straightforwardly be expanded further in $\alpha_s(\mu_R)$. 
If we keep just the diagonal (quark-to-quark) contributions and their LP and lowest-order NLP parts, we
recover the NLO and NNLO terms already contained in Eq.~(\ref{evolnew}). 

In the spin-dependent case the spacelike matrix in Eq.~(\ref{cast}) is to be replaced by the polarized one, 
$\Delta\mathcal{P}_s^N(\alpha_{s})$, given to NLO in~\cite{Mertig:1995ny,Vogelsang:1995vh,Vogelsang:1996im}. 
The helicity evolution kernels $\Delta\mathcal{P}_s^N(\alpha_{s})$ are identical to the unpolarized ones
in the large-$N$ limit at LP. This equality extends even to the first NLP ($1/N$) corrections, except for a difference
$\propto \ln(N)/N$ in the NLO $gq$ splitting function~\cite{Moch:2014sna}. This difference, however, does not affect
the dominant NLP corrections for SIDIS at NNLO. We thus conclude that the approximate NNLO corrections
to be presented next apply to both the spin-averaged and the spin-dependent hard-scattering functions.

\section{Expansion to NNLO \label{NNLOexp}}

We are now ready to present the NNLO (${\cal O}(\alpha_s^2)$) expansion for the SIDIS quark hard-scattering function near
threshold, which is the main result of this paper. We insert the NLP evolved parton distributions and fragmentation 
functions of Eq.~(\ref{evolnew}) into Eq.~(\ref{SIDISres4}) and expand. To write our formulas compactly,
we introduce
\beq
{\cal L}\,\equiv\,\frac{1}{2}\left( \ln (\bar{N})  + \ln (\bar{M})\right) \,.
\eeq 
We then find for the transverse hard-scattering function in the quark channel:
\beeq
\tilde{\omega}_{qq}^{T}\left(N,M,\alpha_{s}(\mu_{R}), \frac{\mu_{R}}{Q}, \frac{\mu_{F}}{Q}\right)&=&
1+\frac{\alpha_s(\mu_R)}{\pi}\,\tilde{\omega}^{T,(1)}_{qq}+\left(\frac{\alpha_s(\mu_R^2)}{\pi}\right)^2 \tilde{\omega}^{T,(2)}_{qq}+
{\cal O}(\alpha_s^3)\,,
\eeeq
where
\beeq\label{eqnlo1}
\tilde{\omega}^{T,(1)}_{qq}\left(N,M, \frac{\mu_{R}}{Q}, \frac{\mu_{F}}{Q}\right)&=&
e^{2}_{q}C_{F}\left\{2{\cal L}^2+\frac{\pi ^2}{6}-4+\left(-\frac{3}{2}+2{\cal L}\right)\ln \frac{\mu_{F}^{2}}{Q^{2}}+ 
{\cal L}\left(\frac{1}{N}+\frac{1}{M}\right)\right\}\,.\quad
\eeeq
The last term is the NLP contribution. We have kept
``mixed'' NLP corrections of the form $\ln(\bar{N})/M$ and $\ln(\bar{M})/N$.
Equation~(\ref{eqnlo1}) reproduces the dominant part of the  full NLO results given in~\cite{Stratmann:2001pb,Anderle:2012rq}, including 
the NLP terms. Its LP part is consistent with the results based on NLL threshold resummation presented in~\cite{Anderle:2012rq}. 

For the approximate NNLO terms we find 
\beeq \label{eqnlo1a}
&&\hspace*{-1.5cm}\frac{1}{e_q^2}\,\tilde{\omega}^{T,(2)}_{qq}\left(N,M, \frac{\mu_{R}}{Q}, \frac{\mu_{F}}{Q}\right)\,=\,
2C_F^2 {\cal L}^4 + 4C_F {\cal L}^3 \left(\frac{\pi}{3} \,b_0+ C_F\ln\frac{\mu^{2}_{F}}{Q^{2}}\right)\nn\\[2mm]
&&+\,C_F{\cal L}^2\left[ C_F \left(-8+\frac{\pi^2}{3}+2 \ln^2\frac{\mu^{2}_{F}}{Q^{2}}-
3 \ln\frac{\mu^{2}_{F}}{Q^{2}}\right)+\left(\frac{67}{18}-\frac{\pi ^2}{6}\right) C_{A}-\frac{5}{9} N_{f}\right]\nn\\[2mm]
&&+\,C_F {\cal L}\left[ \left(\frac{101}{27}-\frac{7}{2}\,\zeta(3)\right)\,C_A -\frac{14}{27}\,N_f+
C_F \ln\frac{\mu^{2}_{F}}{Q^{2}} \left( -8+\frac{\pi^2}{3}-3\ln\frac{\mu^{2}_{F}}{Q^{2}}\right)\right.\nn\\[2mm]
&&\left.+\,\left( \left(\frac{67}{18}-\frac{\pi ^2}{6}\right) C_{A}-\frac{5}{9} N_{f}\right)  \ln\frac{\mu^{2}_{F}}{Q^{2}}
-\pi b_0 \ln^2\frac{\mu^{2}_{F}}{Q^{2}}\right]\nn\\[2mm]
&&+\,C_F^2\left[ \frac{511}{64}-\frac{\pi^2}{16}-\frac{\pi^4}{60}-\frac{15}{4}\,\zeta(3)+
\ln\frac{\mu^{2}_{F}}{Q^{2}}\left( \frac{9}{8}\ln\frac{\mu^{2}_{F}}{Q^{2}}+\frac{93}{16}-3 \zeta(3)\right)\right]  \nn\\[2mm]
&&+\,C_F C_A \left[ -\frac{1535}{192}-\frac{5\pi^2}{16}+\frac{7\pi^4}{720}+\frac{151}{36}\,\zeta(3)\right]+
C_F N_f\left[ \frac{127}{96}+\frac{\pi^2}{24}+\frac{\zeta(3)}{18}\right]  \nn\\[2mm]
&&+\,\frac{3}{4}\,C_F \pi b_0 \ln^2\frac{\mu^{2}_{F}}{Q^{2}}-\frac{C_F \pi^3 b_0}{3}\ln\frac{\mu^{2}_{F}}{Q^{2}}+C_F \left(
-\frac{17}{48}\,C_A + \frac{3}{2}\,\zeta(3)C_A+\frac{N_f}{24}\right)\ln\frac{\mu^{2}_{F}}{Q^{2}}\nn\\[2mm]
&&+\,\pi b_0 \ln\frac{\mu^{2}_{R}}{Q^{2}}\;\frac{1}{e_q^2}\,\tilde{\omega}^{T,(1)}_{qq}\left(N,M, \frac{\mu_{R}}{Q}, \frac{\mu_{F}}{Q}\right)\,
+\,2\,C_F^2\, {\cal L}^3 \left(\frac{1}{N}+\frac{1}{M}\right)\,.
\eeeq
Again, the last term is the dominant NLP correction. Here, two of the three powers of ${\cal L}$ arise from the LP part in the 
first line of Eq.~(\ref{eqnlo1}), which then multiplies the NLO expansion of the NLP factor given
in Eq.~(\ref{nlpfactors}). 

The results for the spin-dependent quark hard-scattering function near threshold are identical:
\beq
\Delta\tilde{\omega}^{(k)}_{qq}\left(N,M, \frac{\mu_{R}}{Q}, \frac{\mu_{F}}{Q}\right)\,=\,
\tilde{\omega}^{T,(k)}_{qq}\left(N,M, \frac{\mu_{R}}{Q}, \frac{\mu_{F}}{Q}\right)\,,
\eeq
for $k=0,1,2$ and including NLP corrections. In fact, this will arguably hold to all orders of perturbation theory. 

So far we have only addressed the $q\to q$ channel. As we discussed in the previous section, the off-diagonal evolution
of parton distributions and fragmentation functions to NLP induces quark-gluon mixing. As a result, once we
insert the NLP singlet evolution in~(\ref{evolit}) into the cross section~(\ref{SIDISres4}), we also obtain terms
with $\tilde{q}(N,\mu_F)\tilde{D}_g(M,\mu_F)$ or $\tilde{g}(N,\mu_F)\tilde{D}_q(M,\mu_F)$.
Evidently, these approximate the quark-to-gluon and gluon-to-quark channel contributions to SIDIS. 
The terms are of course suppressed by $1/N$ or $1/M$, but they also carry logarithmic enhancement. 
We find, at NLO:
\beeq\label{eqnlo2}
\tilde{\omega}^{T,(1)}_{gq}\left(N,M, \frac{\mu_{R}}{Q}, \frac{\mu_{F}}{Q}\right)&=&
-e^{2}_{q}\,C_{F}\,\frac{{\cal L}}{M}\,,\nn\\[2mm]
\tilde{\omega}^{T,(1)}_{qg}\left(N,M, \frac{\mu_{R}}{Q}, \frac{\mu_{F}}{Q}\right)&=&
-e^{2}_{q}\,T_R\, \frac{{\cal L}}{N}\,,
\eeeq
with $T_R=1/2$. These expressions reproduce the corresponding full NLO transverse hard-scattering
functions of Refs.~\cite{Stratmann:2001pb,Anderle:2012rq} at large moment variable. Again the
contributions to the respective spin-dependent hard-scattering functions $\Delta\tilde{\omega}^{(1)}_{gq},
\Delta\tilde{\omega}^{(1)}_{qg}$ are identical to the ones given in~(\ref{eqnlo2}). 

Unfortunately, the evolution method that we have used here to obtain the NLP corrections fails
for the $q\to g$ and $g\to q$ channels beyond NLO. We have found this by inspecting related
results for the Drell-Yan process at measured rapidity. Here, evolution gives the approximate result
\beeq\label{eqnnlo}
\tilde{\omega}^{{\mathrm{DY}},(2)}_{qg}(N,M)\Big|_{\mathrm{evol}}&=&-\frac{T_R{\cal L}}{2M}
\left(4C_F {\cal L}^2 -(C_F-C_A) \ln\bar{N}\ln\bar{M}\right)\,,
\eeeq
whereas the correct result is 
known to be~\cite{Anastasiou:2003ds,Anastasiou:2003yy,Hamberg:1990np,Ravindran:2006bu,Catani:2009sm,Gavin:2010az,Chen:2021vtu}
\beeq\label{eqnnlo1}
\tilde{\omega}^{{\mathrm{DY}},(2)}_{qg}(N,M)&=&-\frac{T_R{\cal L}}{2M}
\left(4C_F {\cal L}^2 -(C_F-C_A) \ln\bar{N}\ln\bar{M}\right)
+(C_F-C_A) \,\frac{\ln^3\bar{M}}{48M}\,.
\eeeq
The difference of the two results is 
\beeq\label{eqnnlo2}
\tilde{\omega}^{{\mathrm{DY}},(2)}_{qg}(N,M)-\tilde{\omega}^{{\mathrm{DY}},(2)}_{qg}(N,M)\Big|_{\mathrm{evol}}
&=&(C_F-C_A)\,
\frac{\ln^3(\bar{M})}{48M}\,.
\eeeq
Interestingly, it depends only on one of the two Mellin variables. In the inclusive case, where $N=M$,
this difference may be understood from Ref.~\cite{LoPresti:2014ihe} where the all-order resummation of the leading
large-$N$ contributions to the quark-gluon contribution to inclusive Drell-Yan was derived. In the light of this, it is
clear that evolution cannot correctly produce the leading NNLO terms for the SIDIS $q\to g$ and $g\to q$ channels.
When expanding our corresponding results, we obtain
\beeq\label{eqnnlo3}
\tilde{\omega}^{T,(2)}_{gq}\left(N,M\right)\Big|_{\mathrm{evol}}&=&-\frac{C_F {\cal L}}{2M}\,
\left(4 C_F {\cal L}^2 - (C_F-C_A) \ln (\bar{N})\ln(\bar{M})\right)\,,\nn\\[2mm]
\tilde{\omega}^{T,(2)}_{qg}\left(N,M\right)\Big|_{\mathrm{evol}}&=&-\frac{T_R {\cal L}}{2N}\,
\left(4C_F {\cal L}^2 - (C_F-C_A) \ln (\bar{N})\ln(\bar{M})\right)\,.
\eeeq
We note that the $1/N$ and $1/M$ terms in the NLO splitting functions contribute here. 
As already stated, the results in Eq.~(\ref{eqnnlo3}) are not expected to be complete, although it appears
likely that a term identical to the one given in Eq.~(\ref{eqnnlo2}) (or with $M\to N$) 
would need to be added. It would be highly desirable
to extend the work of~\cite{LoPresti:2014ihe} to the Drell-Yan process at measured rapidity and to SIDIS.
This is of course beyond the scope of the present work.  For now we therefore refrain from
encouraging use of Eq.~(\ref{eqnnlo3}) in any phenomenological analysis. Our NNLO approximations
given in this paper therefore only apply to the quark channel. 

Appendix~\ref{sec:appendixxzspace} presents our NLO and NNLO near-threshold results as functions of 
$\hat{x}$ and $\hat{z}$. These are obtained by a straightforward inverse transform of the above
Mellin-space results.

\section{Phenomenological predictions \label{Pheno}}

We now turn to a few illustrative phenomenological applications of our approximate NNLO results. Here we only consider
the unpolarized transverse structure function. We reserve a more detailed numerical analysis to future work~\cite{prep},
in which we will also investigate the phenomenology of NNLL resummation.

We first need to go back from Mellin space to $x,z$ space. This is achieved by an inverse
double-Mellin transform. The structure function ${\cal F}^h_i(x,z,Q^2)$ can be recovered
from its moments $\tilde{{\cal F}}^h_i(N,M,Q^2)$ given in Eq.~(\ref{moms}) in the following
way:
\begin{equation}\label{eq:inverse}
{\cal F}^h_i(x,z,Q^2)\,=\, \int_{{\cal C}_N}
\frac{d N}{2\pi i}\,x^{-N}  \int_{{\cal C}_M}
\frac{d M}{2\pi i}\,z^{-M}\,
\tilde{{\cal F}}^h_i(N,M,Q^2)\,,
\end{equation}
where ${\cal C}_N$ and ${\cal C}_M$ denote integration contours in the complex plane, one for each Mellin inverse. 
They have to be chosen in such a way that all singularities of the integrand in $N$ lie to the left of ${{\cal C}_N}$,
and likewise for the poles in $M$ and the contour ${{\cal C}_M}$. In the actual calculation, we obtained excellent 
numerical convergence by setting $N=c_N+\zeta\, {\mathrm{e}}^{i \phi_N}$ and $M=c_M+\xi \,{\mathrm{e}}^{i \phi_M}$
(with $\zeta,\xi\in [0,\infty]$ as contour parameters), where 
$c_N=1.8$ and $c_M=3.3$ and where we tilt each contour by an angle $\phi_N=\phi_M=3\pi/4$. 

To be consistent with the NNLO approximation of the hard-scattering functions that we make, we also need to use NNLO parton distribution
functions and fragmentation functions. For the former, we choose the CT18 NNLO set of Ref.~\cite{Hou:2019efy}, from which we 
also adopt the NNLO strong coupling. NNLO analyses
of fragmentation functions are still scarce~\cite{Anderle:2016czy,Abdolmaleki:2021yjf}, partly because only the process
$e^+e^-\to h+X$ is available at NNLO. For the present study we use the set of Ref.~\cite{Anderle:2016czy}. 
In order to be able to examine the sizes of the various corrections to the cross section, we stick to the
NNLO sets of parton distributions and fragmentation functions also when computing LO or NLO
results. Unless stated otherwise, we choose the renormalization and factorization scales as $\mu_R=\mu_F=Q$. 
Technically, in order to obtain Mellin moments of the parton distributions and fragmentation functions
as needed for Eq.~(\ref{Melmo}) in~(\ref{eq:inverse}), we perform fits of a functional form $P(x)$ to them,
so that the Mellin moments of $P(x)$ can be taken analytically. We have checked that our fits are accurate
to better than $1\%$ over the kinematic domain we are interested in. 

We present results appropriate for the COMPASS experiment at CERN with c.m. energy $\sqrt{s} = 17.3$~GeV,
and for the EIC with $\sqrt{s} = 100$~GeV. For both, we consider the process $\ell p\to \ell\pi^+X$. 
We compute the contribution by the transverse structure function to the SIDIS
cross section, using Eq.~(\ref{sidiscrsec}) and dropping the longitudinal part. We focus on the $z$-dependence of 
the cross section and integrate over $y \in[0.1,0.9]$ and $x \in[0.1,0.8]$. Note that we choose both $x$ and $z$ 
to be rather large so that we are safely in the threshold regime. Because of the relation $Q^2=xys$, 
our choice of kinematics implies $Q^2>3$~GeV$^2$ for COMPASS, and 
$Q^2>100$~GeV$^2$ for the EIC. We furthermore require $W>7$~GeV, where $W^2=Q^2(1-x)/x+m_p^2$.

We note in passing that SIDIS experiments typically quote hadron multiplicities, which are ratios of the SIDIS cross section over
the fully inclusive DIS one, for  given kinematics. For the present paper we are interested in the actual
NNLO corrections to SIDIS, so we do not compute multiplicities here. It would be straightforward to 
do this by computing DIS to full NNLO.  

\begin{figure}
\includegraphics[scale=0.9]{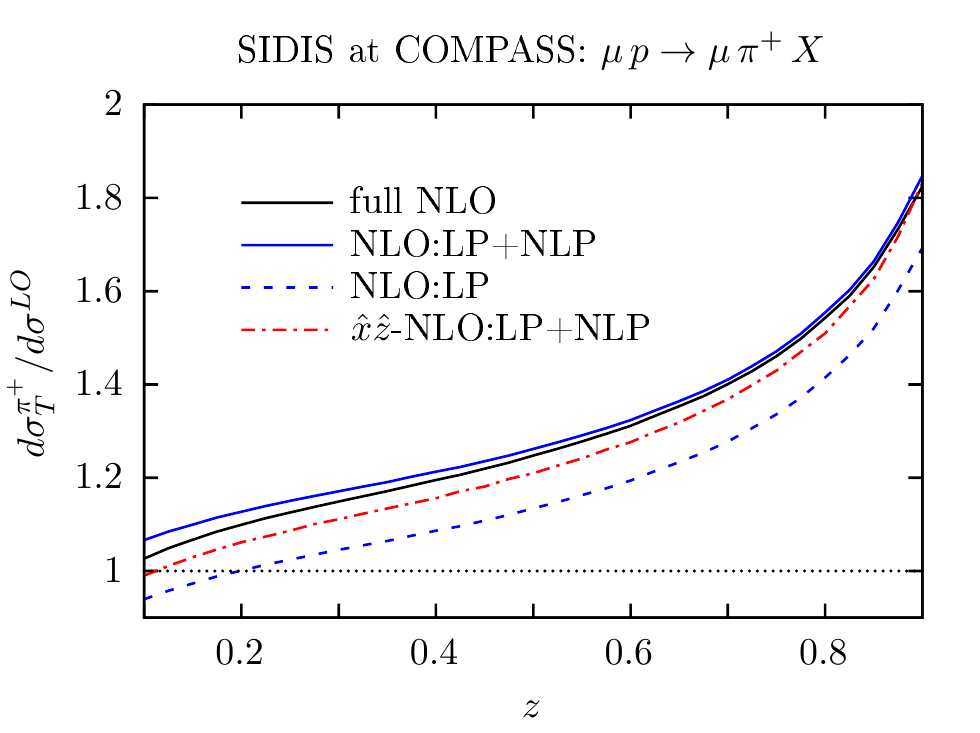}

\vspace*{-6.75cm}
\hspace*{8.6cm}
\includegraphics[scale=0.9]{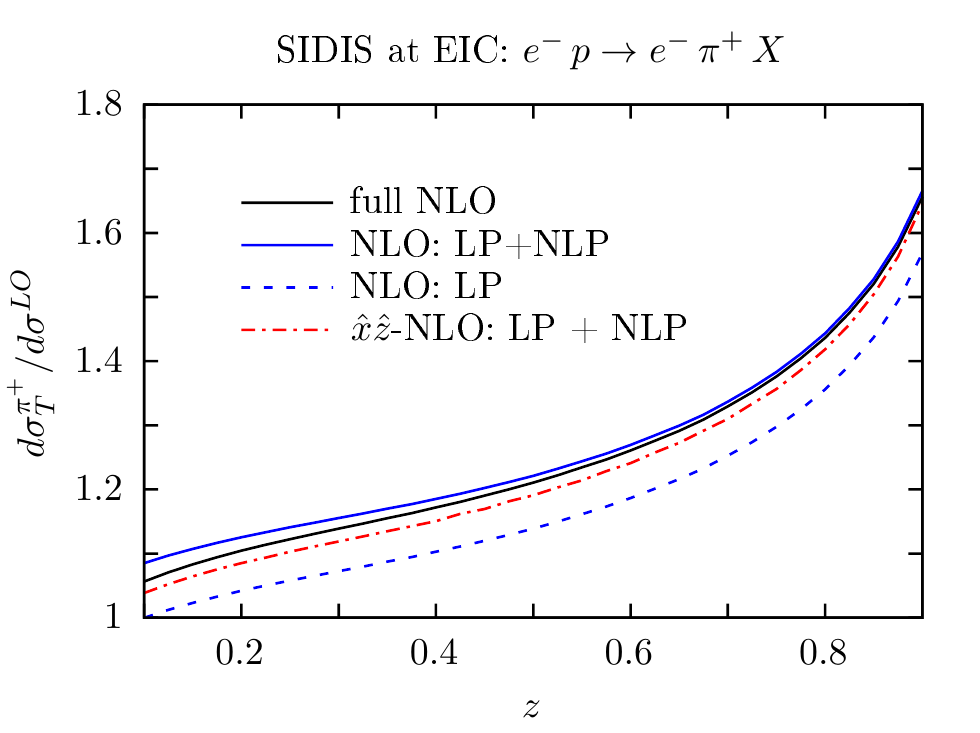}

\caption{{\it Left: Ratios of NLO results for the unpolarized $\ell p\to \ell\pi^+X$ transverse cross section in the $q\to q$ channel to the
LO cross section, for COMPASS kinematics with $x\in [0.1,0.8]$.  The black solid line shows the exact NLO 
result from Ref.~\cite{Stratmann:2001pb}, the dashed blue line the LP approximation in Mellin space, and the solid blue line the 
LP+NLP approximation.  The red dash-dotted line shows the approximation obtained by Eq.~(\ref{Dqq1}) in $\hat{x},\hat{z}$ space.
Right: Same for EIC kinematics.}
\label{fig1}}
\end{figure}

We start by examining NLO, where the exact answer is of course known. In the following we normalize 
all results by the LO cross section. The left part of Fig.~\ref{fig1} presents results for COMPASS
kinematics. The black line shows the ratio of the full transverse
NLO cross section for the $q\to q$ channel to the LO one. As one can see, the NLO corrections 
show the expected strong increase toward large values of $z$. The dashed blue line shows 
the LP approximation to the NLO cross section, based on Eq.~(\ref{eqnlo1}) but without the NLP 
term in the second line. The result shows overall good agreement with full NLO, indicating the 
dominance of the threshold regime, but has a nearly constant difference to the exact result. 
The agreement with full NLO becomes even much better when the dominant
NLP corrections in the second line of Eq.~(\ref{eqnlo1}) are included, as shown by the solid blue line. 
Clearly the full NLO is excellently approximated by this near-threshold result over the whole range in $z$,
and especially so toward large $z$. 

It is interesting to compare the NLO approximations based on the Mellin-space calculation 
(as shown so far) and on Eq.~(\ref{Dqq1}) in $\hat{x},\hat{z}$ space. The two approximations
differ by terms that are even more suppressed than the NLP terms. Nevertheless, their
numerical difference is quite large, with the Mellin result yielding a far better approximation
to the exact NLO result than the approximate $\hat{x},\hat{z}$ space result. We thus
conclude that Mellin space appears better suited for obtaining accurate approximations
to the full result. Similar conclusions were obtained for other processes, such as for Higgs boson production~\cite{Catani:2003zt}.

The right part of Fig.~\ref{fig1} shows corresponding results for the EIC. They have a very similar
trend as our COMPASS results, with a slightly reduced size of the corrections near threshold.
This is expected due to the larger $Q^2$ relevant at the EIC. Again, the NLO corrections are
extremely well reproduced by the approximate ones generated by Mellin-space LP+NLP
resummation. 

\begin{figure}
\includegraphics[scale=0.9]{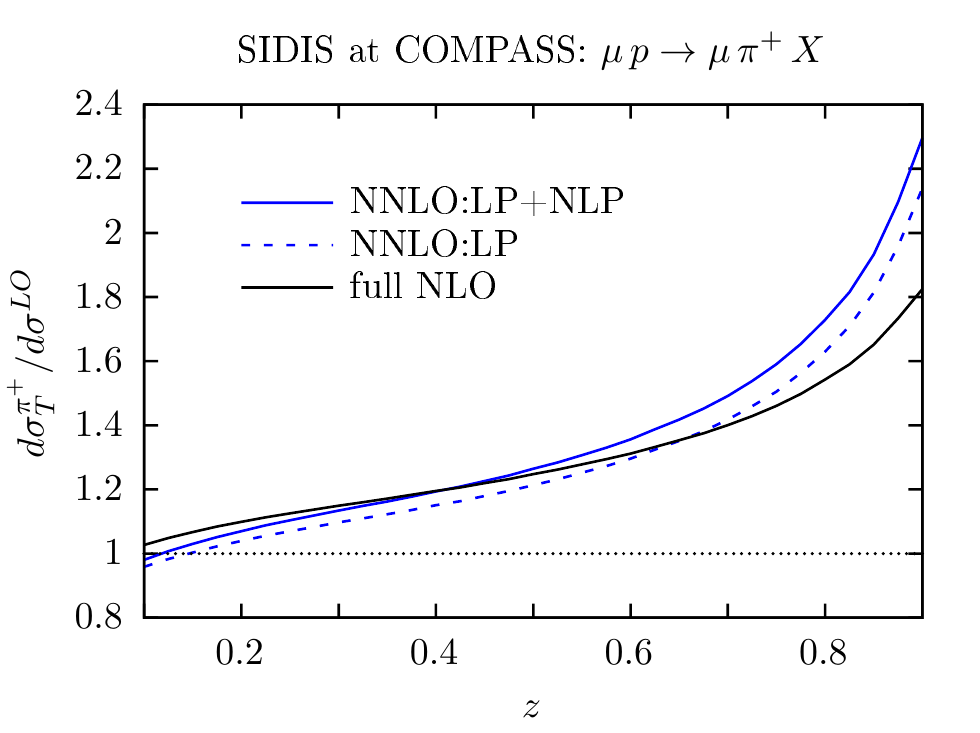}

\vspace*{-6.75cm}
\hspace*{8.6cm}
\includegraphics[scale=0.9]{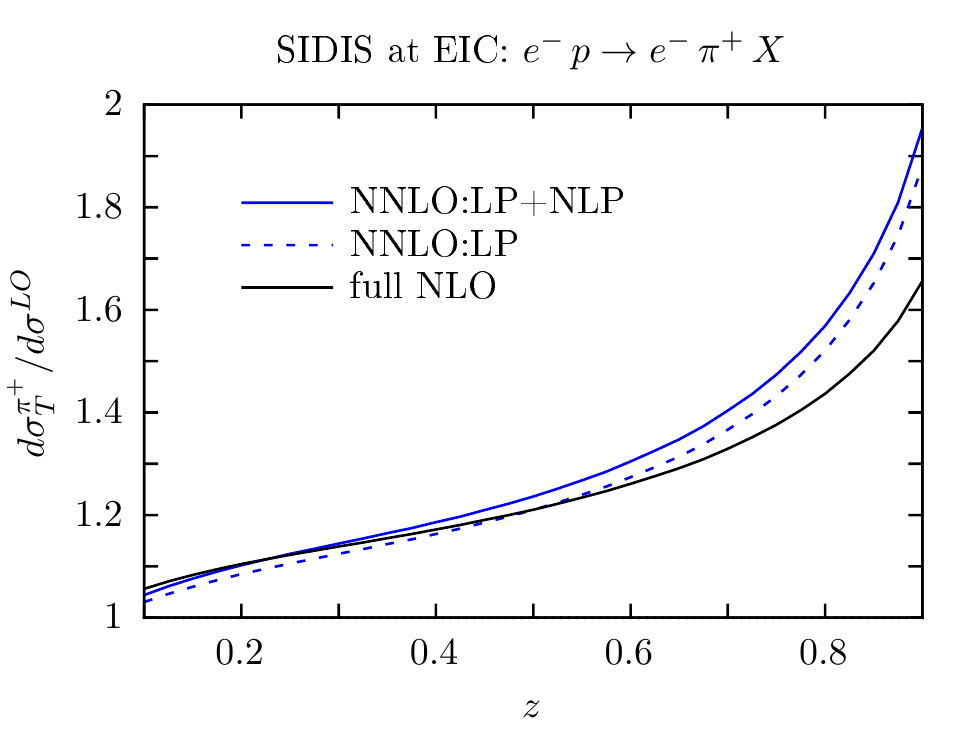}

\caption{{\it Left: Ratios of Mellin-space NNLO results for the unpolarized transverse cross section in the $q\to q$ channel to the
LO cross section, with NLP corrections (blue solid) and without NLPs (blue dashed), for $\ell p\to \ell\pi^+X$ at COMPASS. 
For comparison we also show the full NLO result again. Right: Same for EIC kinematics.}
\label{fig2}}
\end{figure}

The findings in Fig.~\ref{fig1} provide confidence that our Mellin-space NNLO expansions based on resummation 
also provide an accurate approximation to the full NNLO corrections for the $q\to q$ channel. 
Figure~\ref{fig2} presents our NNLO results, again normalized to LO. Here we have included the 
exact NLO part of the cross section, so that the approximation only applies to the NNLO terms. 
The dashed line shows the result based on the LP terms at NNLO, while for the solid one 
we have included the dominant NLP terms as well.  We also display again the curves for full NLO 
that were already shown in Fig.~\ref{fig1}. One can see that the NNLO corrections become
sizable as $z\to 1$, where the threshold logarithms grow in size. As in the NLO case, there
is a rather significant positive contribution to the cross section by the NLP terms, both for 
COMPASS and the EIC.

\begin{figure}
\includegraphics[scale=0.9]{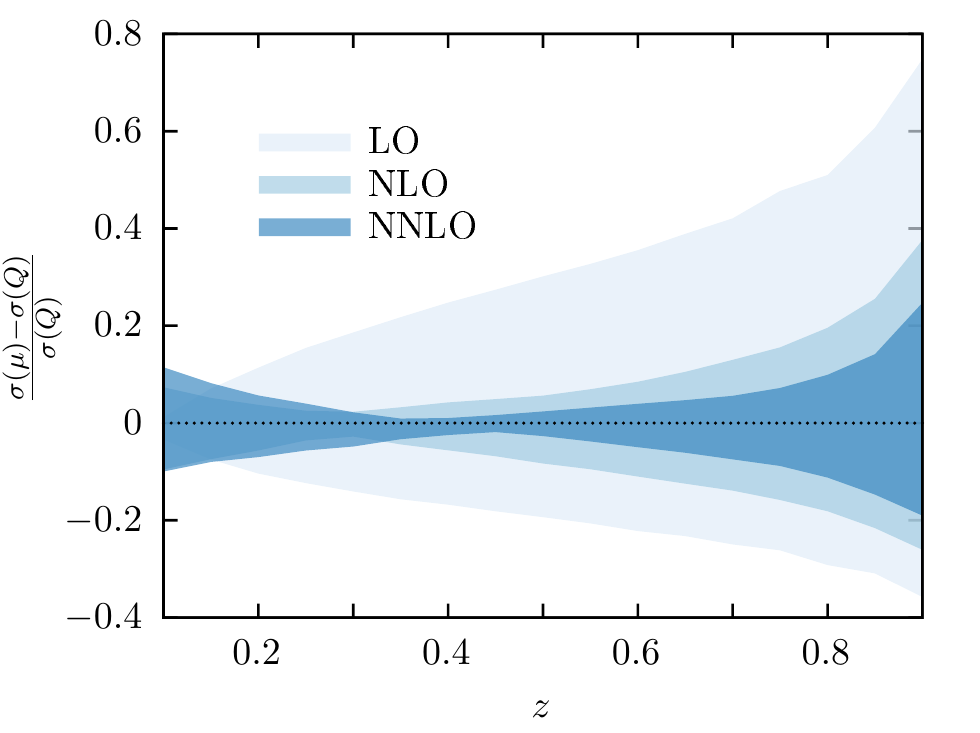}

\vspace*{-6.75cm}
\hspace*{8.6cm}
\includegraphics[scale=0.9]{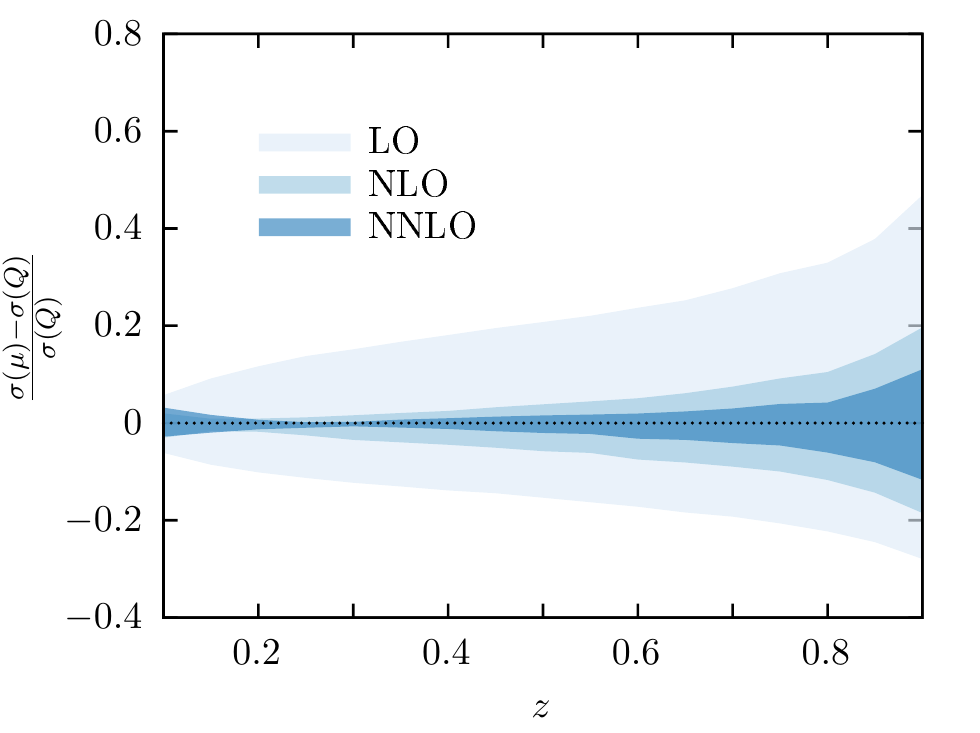}

\caption{{\it Left: Scale dependence of the NLO and approximate NNLO cross sections for COMPASS kinematics. We have varied $\mu_F$
and $\mu_R$ as described in the text. We have used $Q^2>5$~GeV$^2$ here. Right: Same for EIC kinematics.}
\label{fig5}}
\end{figure}

Inclusion of the dominant NNLO terms is expected to reduce the dependence of the cross section on 
the renormalization and factorization scales. Figure~\ref{fig5} shows the variation 
of the LO, full NLO and the (approximate) NNLO cross sections with scale. Here we vary
independently $\mu_F=Q/2,Q,2Q$ and $\mu_R=Q/2,Q,2Q$. Among the nine combinations
this results in, we discard the two with very disparate values, that is, $\mu_F=Q/2,\mu_R=2Q$
and $\mu_F=2Q,\mu_R=Q/2$. We then take the envelope of the remaining seven results.
The figure shows the resulting bands. We present them 
in terms of the ratio $\big(d\sigma(\mu_F,\mu_R)-d\sigma(\mu_F=\mu_R=Q)\big)/d\sigma(\mu_F=\mu_R=Q)$, so that
the cross section with $\mu_F=\mu_R=Q$ always produces the zero line in the plot.
The result for COMPASS\footnote{For COMPASS, we now increase the lower cut on $Q^2$
to $Q^2>5$~GeV$^2$, so that we can reasonably use the scale $\mu=Q/2$.} (left figure)
shows that around $z=0.1$ the NNLO scale 
uncertainty is large, but does improve significantly
toward higher $z$ where it becomes better than the NLO one.
It does, however, remain non-negligible even at large $z$. The main patterns are reproduced also 
for EIC kinematics (right part of the figure); however, here the scale uncertainty is overall very small
at NNLO, showing a band that is much narrower than the NLO one at medium to large $z$. 
We note that we have included the NLO contributions by the $qg$ subprocesses in the
results shown in the figure, whose effects are however relatively small. 

We finally note that we do not consider the SIDIS spin asymmetry here. Since the approximate 
NNLO corrections are identical for the spin-averaged and spin-dependent cross sections 
(even for the dominant NLP terms), the asymmetry is expected to be affected very little
by the corrections. This was indeed already observed in the NLL study~\cite{Anderle:2013lka}.

\section{Conclusions and outlook}

We have presented approximate next-to-next-to-leading order corrections to semi-inclusive DIS,
$\ell p\to \ell hX$. These corrections apply to the quark channel and are based on the threshold resummation
formalism. We have first determined all ingredients for threshold resummation for SIDIS at next-to-next-to-leading
logarithmic accuracy, extending previous work by one logarithmic order. As SIDIS is characterized
by two ``scaling'' variables, $\hat{x}=Q^{2}/2 p \cdot q$ and $\hat{z}=p\cdot p_c/P \cdot q$,
the moment-space resummation is naturally formulated in terms of two Mellin moments $N$ and $M$.
Although these are separate variables, the SIDIS resummation formula may be obtained by that for 
the inclusive Drell-Yan cross section by a simple rescaling $N\to \sqrt{NM}$, up to differences associated
with the fact that the virtual photon in SIDIS is spacelike. These differences are accounted for
by the hard factor in the resummed SIDIS cross section, which is related to the spacelike quark
form factor rather than the timelike one contributing to Drell-Yan. We have subsequently expanded
the resummed expressions to ${\mathcal O}(\alpha_s^2)$, to obtain the NNLO corrections. 

To further improve the accuracy of the near-threshold NNLL and NNLO approximation we have determined
also the dominant subleading terms which are suppressed by $1/N$ or $1/M$ near threshold,
but still enhanced by logarithms. These ``next-to-leading power'' terms may be obtained
by use of the DGLAP evolution of the parton distribution functions between scales $Q$ and
$Q/\sqrt{\bar{N}\bar{M}}$. We have found that the approximate NNLO corrections are
identical for the spin-averaged (transverse) cross section and the longitudinally polarized
one, even including the NLP corrections. This has important ramifications for phenomenology
as it means that the SIDIS spin asymmetry will be largely unaffected even by NNLO corrections. 

We have presented a few basic phenomenological results at approximate NNLO. 
These indicate a significant increase of the cross section at large $z$, as well as a still sizable
contribution of the NLP corrections. Our results are readily suited for initial studies
of SIDIS at NNLO in ``global'' fitting frameworks for fragmentation functions and/or
parton distributions, especially polarized ones. Furthermore, the corrections we have
derived will provide important benchmarks for future full NNLO calculations of SIDIS. 

There are several avenues for future improvements on our work. Extension to 
approximate N$^3$LO near threshold and to N$^3$LL resummation would be 
quite straightforward~\cite{prep}. As already mentioned earlier, it will also be important to address the quark-gluon channels to SIDIS
and to determine their dominant NLP corrections, following the lines in Ref.~\cite{LoPresti:2014ihe}.
In the same vein, the longitudinal SIDIS structure function should be addressed at higher orders.
For inclusive DIS, $F_L$ receives corrections as large as $\alpha_s^2\ln^2(1-x)$ at high $x$, which 
were derived and extended to all orders in Ref.~\cite{Moch:2009mu}. Although these are again NLP
corrections, it will be relevant to investigate the corresponding logarithmic structure of $F_L$ in SIDIS.
Finally, we note that the corrections we have derived here are really valid when both $x$ and $z$
are large. The recent study~\cite{Lustermans:2019cau} considers the Drell-Yan cross section
at measured rapidity and derives a factorization theorem that is valid when only {\it one} of the
two kinematic variables $\sqrt{z}\,{\mathrm{e}}^{\pm y}$ is large, while the other can have an arbitrary
value. Extension of such a theorem to the SIDIS case when only $x$ {\it or} $z$ is large would
be quite valuable as it would extend the validity of the threshold approximation for SIDIS.

\section*{Acknowledgements}

We acknowledge useful communications with F. Olness, V. Ravindran, O. Sch\"{u}le, G. Sterman, and 
M. Stratmann. We thank V. Bertone for pointing us to the numerical code for Ref.~\cite{Anastasiou:2003ds},
from which the the expression in Eq.~(\ref{eqnnlo1}) was extracted.
This study was supported in part by Deutsche Forschungsgemeinschaft (DFG) through the 
Research Unit FOR 2926 (project number 40824754).

\appendix
\section{Appendix: Coefficients for resummation to NNLL}
\label{sec:appendixAndim}

We use the following expansion of the running strong coupling~\cite{Vogt:2000ci,Catani:2003zt}
\beeq
\alpha_{s}(\mu)
	&=&\frac{\alpha_{s}(\mu_{R}) }{X}\left[1
		-\frac{\alpha_{s}(\mu_{R}) }{ X} \frac{b_{1}}{b_{0}} \ln X\right.\nn\\[2mm]
	&+&\left.\left(\frac{\alpha_{s}(\mu_{R})}{X}\right)^{2} \left(\frac{b_{1}^2}{b_{0}^2 } \left(\ln ^2 X-\ln X+X-1\right)-\frac{b_{2}}{b_{0}} (X-1)\right)\right]\,,
\label{eq:asmu}
\eeeq
where
\begin{equation}
X\,\equiv \,1 + b_{0} \alpha_{s} (\mu_{R}) \ln \frac{\mu^{2}}{\mu^{2}_{R}}\,,
\end{equation}
and
\bea
b_0 & = & \f{1}{12\pi} \left(11C_A-2 N_f\right)\; , \qquad b_1 = \f{1}{24\pi^2}\left(17C_A^2-5C_AN_f-3C_F N_f\right) \; , \nn \\[2mm]
b_2 & = & \f{1}{64\pi^3}\left(\f{2857}{54} C_A^3- \f{1415}{54} C_A^2 N_f-\f{205}{18} C_A C_F N_f+C_F^2 N_f+
\f{79}{54} C_A N_f^2+\f{11}{9} C_F N_f^2\right) ,\;\;
\eea
with $N_f$ the number of flavors and 
\beq\label{cqcg} 
C_F\,=\,\frac{N_c^2-1}{2N_c}\,=\,\frac{4}{3}  
\;, \;\;\;C_A\,=\,N_c=3 \; .
\eeq
The various functions we use in the main text have the following perturbative expansions:
\bea
A_q(\as) & = & \frac{\as}{\pi} \,A_q^{(1)} +\left( \frac{\as}{\pi}\right)^2 A_q^{(2)} + 
\left(\frac{\as}{\pi}\right)^3 A_q^{(3)}+{\mathcal O}(\as^4) \,,\nn \\[2mm]
\widehat{D}_q(\as) & = & \left(\f{\as}{\pi}\right)^2 \widehat{D}_q^{(2)}+{\mathcal O}(\as^3) \,,\nn\\[2mm]
P_{q,\delta}(\as) &=& \frac{\as}{\pi} \,P_{q,\delta}^{(1)} +\left( \frac{\as}{\pi}\right)^2 P_{q,\delta}^{(2)} + 
{\mathcal O}(\as^3)\,,
\eea
where to NNLL we use the coefficients of $A_q$ from~\cite{Vogt:2000ci,KT,Moch:2004pa,Harlander:2001is,eric}:
\beeq 
\label{A12coef} 
A_q^{(1)} &=&C_F
\;,\;\;\;\;\quad A_q^{(2)}\;=\;\frac{1}{2} \; C_F \left[ 
C_A \left( \frac{67}{18} - \frac{\pi^2}{6} \right)  
- \frac{5}{9} N_f \right] \; , \nn \\[2mm] 
A_q^{(3)}&=&\f{1}{4}C_F\left[C_A^2 \left(\f{245}{24}-\f{67}{9}\zeta(2)+
\f{11}{6}\zeta(3)+\f{11}{5}\zeta(2)^2 \right)+C_F N_f\left(-\f{55}{24}+2\zeta(3) \right)\right. \nn \\[2mm]
&& \qquad\;\, \left. +C_A N_f\left(-\f{209}{108}+\f{10}{9}\zeta(2)-\f{7}{3}\zeta(3) \right) -\f{1}{27}N_f^2 \right] \; , 
\eeeq
Furthermore~\cite{Vogt:2000ci,Catani:2001ic,Catani:2003zt,Hinderer:2018nkb},
\bea\label{eq:Dhat2}
\widehat{D}_q^{(2)}
&=&C_F\,\left[C_A\left(-\f{101}{27}+\f{7}{2}\,\zeta(3)\right) + \f{14}{27}\,N_f   \right]\, ,
\eea
and~\cite{Curci:1980uw}
\beeq
P_{q,\delta}^{(1)}&=&\frac{3}{4}\,C_F\,, \nn \\[2mm]
P_{q,\delta}^{(2)}&=&\frac{1}{4}\left[ C_F^2 \left( \frac{3}{8}-3\zeta(2) +6 \zeta(3)\right) + C_F C_A 
\left(\frac{17}{24}+\frac{11}{3}\zeta(2) -3 \zeta(3)\right) -
\frac{C_F N_f}{2} \left(\frac{1}{6}+\frac{4}{3} \zeta(2) \right)\right]\,.\nn\\
\eeeq
Finally, the coefficient $\widehat{C}_{qq}$ in Eq.~(\ref{SIDISres1}) is expanded as
\begin{equation}
\widehat{C}_{qq} \left(\alpha_{s}\big(\mu_{R}\big),\frac{\mu_{R}}{Q}
\right)		\,=\, 1 + \frac{\alpha_{s}\big(\mu_{R}\big)}{\pi}\, \widehat{C}_{qq}^{(1)}
			+ \left( \frac{\alpha_{s}(\mu_{R}\big)}{\pi} \right)^{2} \widehat{C}_{qq}^{(2)}
			+ \mathcal{O}(\alpha^{3}_{s})\,,
\end{equation}
with~\cite{Catani:2003zt}
\beeq
\widehat{C}_{qq}^{(1)}
	&=& \frac{\pi ^2 }{3}A^{(1)}_{q}\,,\\[2mm]
\widehat{C}_{qq}^{(2)}
	&=&\frac{\pi ^4 }{18}\big(A^{(1)}_q\big)^2
		+A^{(1)}_{q} \pi b_{0} \left(\frac{\pi ^2}{3}  \ln \frac{\mu^{2}_{R}}{Q^{2}} 
		+\frac{8}{3} \zeta (3)  \right) 
		+\frac{\pi ^2 }{3}A^{(2)}_{q}\,.
\eeeq

\section{Appendix: Near-threshold results in $x,z$-space}
\label{sec:appendixxzspace}

Performing a double-inverse Mellin transform of the results given in Eqs.~(\ref{eqnlo1}) and~(\ref{eqnlo1a}) 
we obtain approximate results for the quark-to-quark hard-scattering function
$\omega^T_{qq} \left(\hat{x},\hat{z},\alpha_{s}(\mu_{R}), \mu_{R}/Q, \mu_{F}/Q\right)$ in Eq.~(\ref{F1hallorders}), 
valid near threshold. To obtain compact expressions, we introduce the following abbreviations:
\beeq
\delta_x&\equiv&\delta(1-\hat{x})\,, \quad \delta_z\,\equiv\,\delta(1-\hat{z})\,,\nn\\[2mm]
\mathcal{D}_x^{i} &\equiv& \left[\,\frac{\ln^{i}(1-\hat{x})}{1-\hat{x}}\,\right]_{+} \,, \quad 
\mathcal{D}_z^{i} \,\equiv\, \left[\,\frac{\ln^{i}(1-\hat{z})}{1-\hat{z}}\,\right]_{+} \,,\nn\\[2mm]
\ell_x^{i}&\equiv& \ln^{i}(1-\hat{x})\,, \quad\ell_z^{i}\,\equiv\, \ln^{i}(1-\hat{z})\,.
\eeeq
We write 
\beeq
\frac{1}{e_q^2}\,\omega^T_{qq} \left(\hat{x},\hat{z},\alpha_{s}, \frac{\mu_{R}}{Q}, \frac{\mu_{F}}{Q} \right)
	&=& \delta_x\,\delta_z + \frac{\alpha_{s}}{\pi} \; C_F\,\Delta^{(1)}_{qq}\nn \\[2mm]
	&+& \left(\frac{\alpha_{s}}{\pi} \right)^2C_F
		\left[ C_F\,\Delta^{(2), C_{F}}_{qq}
		+  C_A\,\Delta^{(2), C_{A}}_{qq} 
		+   N_f\,\Delta^{(2), N_{f}}_{qq} \right]\nn \\[2mm]
	&+&\left(\frac{\alpha_{s}}{\pi} \right)^2 \pi b_{0} \;  C_F\,\Delta^{(1)}_{qq}\; \ln\frac{\mu^{2}_{R}}{Q^{2}}\,+\,
	{\cal O}(\alpha_s^3)\,.
\eeeq
The NLO term reads
\beeq
\label{Dqq1}
\Delta^{(1)}_{qq} &=&\delta_x \mathcal{D}_z^1+\delta_z \mathcal{D}_x^1+\mathcal{D}_x^0\,\mathcal{D}_z^0
	-4 \delta_x\delta_z -
	\delta_x \left( \mathcal{D}_z^0+\frac{3}{4} \delta_z\right)\ln\frac{\mu^{2}_{F}}{Q^{2}}-
	\delta_z \left( \mathcal{D}_x^0+\frac{3}{4} \delta_x\right)\ln\frac{\mu^{2}_{F}}{Q^{2}}\nn\\[2mm]
	&-&\mathcal{D}_x^0-\mathcal{D}_z^0 - \delta_x \, \ell^1_z - \delta_z\,\ell^1_x  \,.
\eeeq
Note that we have included the NLP contributions, which are given in the second line. They show up as terms that carry only a single distribution,
in either $\hat{x}$ or $\hat{z}$. 

Since the NNLO $C_F^2$ contribution is quite lengthy, we split it into its LP and NLP contributions and write it as
\beq
\Delta^{(2),C_F}_{qq}\,=\,\Delta^{(2),C_F}_{qq,{\mathrm{LP}}}+\Delta^{(2),C_F}_{qq,{\mathrm{NLP}}}\,.
\eeq
We then have for the leading-power part:
\beeq\label{nnlocf2}
\Delta^{(2),C_{F}}_{qq,{\mathrm{LP}}}
	&=& \frac{1}{2}\left(\delta_x\,\mathcal{D}^3_z+\delta_z\,\mathcal{D}^3_x\right)
	+\frac{3}{2}\left(\mathcal{D}^0_x\mathcal{D}^2_z+\mathcal{D}^0_z\mathcal{D}^2_x
+2\, \mathcal{D}^1_x\mathcal{D}^1_z \right)\nn\\[2mm]
&-&\left(4+\frac{\pi ^2 }{3} \right)\left( \mathcal{D}^0_x\mathcal{D}^0_z+
\delta_x\,\mathcal{D}^1_z+\delta_z\mathcal{D}^1_x\right)+
2 \zeta (3)\left( \delta_x\,\mathcal{D}^0_z +\delta_z\,\mathcal{D}^0_x \right)\nn\\[2mm]
&+&\delta_x\,\delta_z\left(\frac{511}{64}-\frac{15 \zeta (3)}{4}+\frac{29 \pi ^2}{48}-\frac{7 \pi ^4}{360}\right)
\nn\\[2mm] 
&+&\left[\delta_x\,\mathcal{D}^1_z+\delta_z\,\mathcal{D}^1_x+\mathcal{D}^0_x \,\mathcal{D}^0_z+
\frac{3}{2}\left(\delta_x\,\mathcal{D}^0_z+\delta_z\,\mathcal{D}^0_x \right)+\delta_x\,\delta_z
\left(\frac{9 }{8}-\frac{\pi ^2}{6}\right)\right]\ln^{2}\frac{\mu^{2}_{F}}{Q^{2}}\nn\\[2mm]
&+&\left[ - \frac{3}{2}\left(\delta_x\,\mathcal{D}^2_z+\delta_z\,\mathcal{D}^2_x+2\, \mathcal{D}^0_x\,\mathcal{D}^1_z+
2\, \mathcal{D}^0_z\,\mathcal{D}^1_x+\mathcal{D}^0_x\,\mathcal{D}^0_z+\delta_x\,\mathcal{D}^1_z
+\delta_z\,\mathcal{D}^1_x\right)\right.\nn\\[2mm]
&&\left. +\left(4+\frac{\pi ^2}{3}\right)\left( \delta_x\,\mathcal{D}^0_z+\delta_z\,\mathcal{D}^0_x\right)+
\delta_x\,\delta_z\left(-5\zeta (3)+\frac{\pi ^2}{4}+\frac{93}{16}\right)
\right] \ln\frac{\mu^{2}_{F}}{Q^{2}}\,,
\eeeq
while the dominant NLP terms are given by
\beeq\label{nnlocf2nlp}
\Delta^{(2),C_{F}}_{qq,{\mathrm{NLP}}}
&=&-\frac{3}{2}\left(  \mathcal{D}^{2}_x+\mathcal{D}^{2}_z+2\,\mathcal{D}^1_x\,\ell^1_z
+2\,\mathcal{D}^1_z\,\ell^1_x+\mathcal{D}^0_x\,\ell^{2}_z+\mathcal{D}^0_z\,\ell^{2}_x\right) - 
\frac{1}{2}\left(\delta_x \,\ell^{3}_z+
		\delta_z \,\ell^{3}_x\right) \,.
\eeeq
The $C_FC_{A}$ and $C_FN_{f}$ parts do not possess any dominant NLP contributions (see Eq.~(\ref{eqnlo1a})).
They read:
\beeq\label{nnlocfca}
\Delta^{(2),C_{A}}_{qq}&=&-\frac{11}{24}\,\left(\delta_x\,\mathcal{D}^{2}_z+\delta_z\,\mathcal{D}^{2}_x
		+2\,\mathcal{D}^0_x\,\mathcal{D}^1_z+2\,\mathcal{D}^0_z\,\mathcal{D}^1_x\right)
+\left(\frac{67 }{36}-\frac{\pi ^2}{12}\right) \left(\mathcal{D}^{0}_x\,\mathcal{D}^{0}_z+\delta_x\,\mathcal{D}^1_z+
\delta_z\,\mathcal{D}^1_x\right)\nn\\[2mm]
&+&\left(\delta_x\,\mathcal{D}^0_z+\delta_z\,\mathcal{D}^0_x\right)\left(\frac{7 \zeta (3)}{4}+\frac{11 \pi ^2}{72}
		-\frac{101}{54} \right)+\delta_x\,\delta_z\left(\frac{43\zeta (3)}{12}
		+\frac{17 \pi ^4}{720}
		-\frac{1535}{192}
		-\frac{269 \pi ^2}{432}\right)\nn\\[2mm]
	&+&\frac{11}{24}\left[ \delta_x\,\mathcal{D}^{0}_z+\delta_z\,\mathcal{D}^{0}_x+
	\frac{3}{2}\,\delta_x\,\delta_z\right]\ln^{2}\frac{\mu^{2}_{F}}{Q^{2}}\nn\\[2mm]
	&+&\left[ -(\delta_x\,\mathcal{D}^{0}_z+\delta_z\,\mathcal{D}^{0}_x)
	\left(\frac{67}{36}-\frac{\pi ^2}{12}\right) +\delta_x\,\delta_z\left(\frac{3 \zeta (3)}{2}-\frac{11 \pi ^2}{36}-\frac{17}{48}\right)
	\right]\ln\frac{\mu^{2}_{F}}{Q^{2}} \,.
\eeeq
and
\beeq\label{nnlocfnf}
\Delta^{(2),N_{f}}_{qq}
	&=&  \frac{1}{12}\,\left(\delta_x\,\mathcal{D}^{2}_z+\delta_z\,\mathcal{D}^{2}_x
		+2\,\mathcal{D}^0_x\,\mathcal{D}^1_z+2\,\mathcal{D}^0_z\,\mathcal{D}^1_x\right)	
		- \frac{5}{18}\,\left(\mathcal{D}^{0}_x\,\mathcal{D}^{0}_z+\delta_x\,\mathcal{D}^1_z+
\delta_z\,\mathcal{D}^1_x\right)\nn\\[2mm]
&+&\left(\delta_x\,\mathcal{D}^0_z+\delta_z\,\mathcal{D}^0_x\right)\left( \frac{7}{27}-\frac{\pi ^2}{36}\right)
+\delta_x\,\delta_z\left(\frac{\zeta (3)}{6}+\frac{19 \pi ^2}{216}+\frac{127}{96}\right) \nn\\[2mm]
	&-& \frac{1}{12}\left[ \delta_x\,\mathcal{D}^{0}_z+\delta_z\,\mathcal{D}^{0}_x+
	\frac{3}{2}\,\delta_x\,\delta_z\right]\ln^{2}\frac{\mu^{2}_{F}}{Q^{2}}\nn\\[2mm]		
	& +&\left[\frac{5}{18}\left( \delta_x\,\mathcal{D}^{0}_z + \delta_z\,\mathcal{D}^{0}_x \right)
+ \delta_x\,\delta_z \left( \frac{1}{24}+\frac{\pi ^2}{18}\right)\right]\ln\frac{\mu^{2}_{F}}{Q^{2}}\,.
\eeeq
We stress that the results in Eqs.~(\ref{nnlocf2}),(\ref{nnlocfca}),(\ref{nnlocfnf}) collect {\it all} double distributions in $\hat{x}$
and $\hat{z}$ that arise at NNLO. We also note that the Mellin-space and the $x,z$-space expressions near
threshold are not strictly identical, but differ by terms that are suppressed near threshold. These terms are generically
of the form $\ln^m(\bar{N})/N^2$ or $1/N$ (without logarithms) in Mellin space (and likewise with $N$ replaced
by $M$), and of the form $(1-\hat{x})\ln^m(1-\hat{x})$ or constant (and also with $\hat{x}$ replaced
by $\hat{z}$) in $x,z$-space.




\newpage
\begin{thebibliography}{99}

\bibitem{deFlorian:2017lwf}
D.~de Florian, M.~Epele, R.~J.~Hernandez-Pinto, R.~Sassot and M.~Stratmann,
Phys. Rev. D \textbf{95}, no.9, 094019 (2017)
[arXiv:1702.06353 [hep-ph]].

\bibitem{Anderle:2016czy}
D.~P.~Anderle, T.~Kaufmann, M.~Stratmann and F.~Ringer,
Phys. Rev. D \textbf{95}, no.5, 054003 (2017)
[arXiv:1611.03371 [hep-ph]].

\bibitem{Leader:2015hna}
E.~Leader, A.~V.~Sidorov and D.~B.~Stamenov,
Phys. Rev. D \textbf{93}, no.7, 074026 (2016)
[arXiv:1506.06381 [hep-ph]].

\bibitem{Bertone:2018ecm}
V.~Bertone \textit{et al.} [NNPDF],
Eur. Phys. J. C \textbf{78}, no.8, 651 (2018)
[arXiv:1807.03310 [hep-ph]].

\bibitem{Khalek:2021gxf}
R.~A.~Khalek, V.~Bertone and E.~R.~Nocera,
Phys. Rev. D \textbf{104}, no.3, 034007 (2021)
[arXiv:2105.08725 [hep-ph]].

\bibitem{deFlorian:2009vb}
D.~de Florian, R.~Sassot, M.~Stratmann and W.~Vogelsang,
Phys. Rev. D \textbf{80}, 034030 (2009)
[arXiv:0904.3821 [hep-ph]];
Phys. Rev. Lett. \textbf{113}, no.1, 012001 (2014)
[arXiv:1404.4293 [hep-ph]].

\bibitem{DeFlorian:2019xxt}
D.~de Florian, G.~A.~Lucero, R.~Sassot, M.~Stratmann and W.~Vogelsang,
Phys. Rev. D \textbf{100}, no.11, 114027 (2019)
[arXiv:1902.10548 [hep-ph]].

\bibitem{Ethier:2017zbq}
J.~J.~Ethier, N.~Sato and W.~Melnitchouk,
Phys. Rev. Lett. \textbf{119}, no.13, 132001 (2017)
[arXiv:1705.05889 [hep-ph]].

\bibitem{Moffat:2021dji}
E.~Moffat \textit{et al.} [Jefferson Lab Angular Momentum (JAM)],
Phys. Rev. D \textbf{104}, no.1, 016015 (2021)
[arXiv:2101.04664 [hep-ph]].

\bibitem{Daleo:2003xg}
A.~Daleo, C.~A.~Garcia Canal and R.~Sassot,
Nucl. Phys. B \textbf{662}, 334 (2003)
[arXiv:hep-ph/0303199 [hep-ph]].

\bibitem{Daleo:2003jf}
A.~Daleo and R.~Sassot,
Nucl. Phys. B \textbf{673}, 357 (2003)
[arXiv:hep-ph/0309073 [hep-ph]].

\bibitem{Anderle:2016kwa}
D.~Anderle, D.~de Florian and Y.~Rotstein Habarnau,
Phys. Rev. D \textbf{95}, no.3, 034027 (2017)
[arXiv:1612.01293 [hep-ph]].

\bibitem{AbdulKhalek:2021gbh}
R.~Abdul Khalek, A.~Accardi, J.~Adam, D.~Adamiak, W.~Akers, M.~Albaladejo, A.~Al-bataineh, M.~G.~Alexeev, F.~Ameli and P.~Antonioli, \textit{et al.}
[arXiv:2103.05419 [physics.ins-det]].

\bibitem{Cacciari:2001cw}
M.~Cacciari and S.~Catani,
Nucl. Phys. B \textbf{617}, 253 (2001)
[arXiv:hep-ph/0107138 [hep-ph]].

\bibitem{Anderle:2012rq}
D.~P.~Anderle, F.~Ringer and W.~Vogelsang,
Phys. Rev. D \textbf{87}, no.3, 034014 (2013)
[arXiv:1212.2099 [hep-ph]].

\bibitem{Anderle:2013lka}
D.~P.~Anderle, F.~Ringer and W.~Vogelsang,
Phys. Rev. D \textbf{87}, 094021 (2013)
[arXiv:1304.1373 [hep-ph]].

\bibitem{Sterman:2006hu}
G.~F.~Sterman and W.~Vogelsang,
Phys. Rev. D \textbf{74}, 114002 (2006)
[arXiv:hep-ph/0606211 [hep-ph]].

\bibitem{Westmark:2017uig}
D.~Westmark and J.~F.~Owens,
Phys. Rev. D \textbf{95}, no.5, 056024 (2017)
[arXiv:1701.06716 [hep-ph]].

\bibitem{Banerjee:2018vvb}
P.~Banerjee, G.~Das, P.~K.~Dhani and V.~Ravindran,
Phys. Rev. D \textbf{98} (2018) no.5, 054018
[arXiv:1805.01186 [hep-ph]].

\bibitem{Catani:2013tia}
S.~Catani, L.~Cieri, D.~de Florian, G.~Ferrera and M.~Grazzini,
Nucl. Phys. B \textbf{881} (2014), 414-443
[arXiv:1311.1654 [hep-ph]].

\bibitem{Catani:2014uta}
S.~Catani, L.~Cieri, D.~de Florian, G.~Ferrera and M.~Grazzini,
Nucl. Phys. B \textbf{888}, 75 (2014)
[arXiv:1405.4827 [hep-ph]].

\bibitem{Gehrmann:2005pd}
T.~Gehrmann, T.~Huber and D.~Maitre,
Phys. Lett. B \textbf{622}, 295 (2005)
[arXiv:hep-ph/0507061 [hep-ph]].

\bibitem{Moch:2005tm}
S.~Moch, J.~A.~M.~Vermaseren and A.~Vogt,
Phys. Lett. B \textbf{625}, 245 (2005)
[arXiv:hep-ph/0508055 [hep-ph]].

\bibitem{deFlorian:1997zj} 
  D.~de Florian, M.~Stratmann and W.~Vogelsang,
  Phys.\ Rev.\ D {\bf 57}, 5811 (1998)
  [hep-ph/9711387].

\bibitem{altarelli} G.\ Altarelli, R.K.\ Ellis, G.\ Martinelli, and S.Y. Pi, 
Nucl. Phys. {\bf B160}, 301 (1979).

\bibitem{Nason:1993xx} 
  P.~Nason and B.~R.~Webber,
  Nucl.\ Phys.\ B {\bf 421}, 473 (1994)
  [Erratum-ibid.\ B {\bf 480}, 755 (1996)].
  
\bibitem{Furmanski:1981cw}
W.~Furmanski and R.~Petronzio,
Z. Phys. C \textbf{11}, 293 (1982).
 
\bibitem{graudenz} D.\ Graudenz, Nucl. Phys. {\bf B432}, 351 (1994).
 
\bibitem{deFlorian:2012wk}
D.~de Florian and Y.~Rotstein Habarnau,
Eur. Phys. J. C \textbf{73} (2013) no.3, 2356
[arXiv:1210.7203 [hep-ph]].

\bibitem{Stratmann:2001pb}
M.~Stratmann and W.~Vogelsang,
Phys. Rev. D \textbf{64}, 114007 (2001)
[arXiv:hep-ph/0107064 [hep-ph]].

\bibitem{Catani:2003zt}
S.~Catani, D.~de Florian, M.~Grazzini and P.~Nason,
JHEP \textbf{07}, 028 (2003)
[arXiv:hep-ph/0306211 [hep-ph]].

\bibitem{Hinderer:2018nkb}
P.~Hinderer, F.~Ringer, G.~Sterman and W.~Vogelsang,
Phys. Rev. D \textbf{99}, no.5, 054019 (2019)
[arXiv:1812.00915 [hep-ph]].

\bibitem{Catani:1989ne}
S.~Catani and L.~Trentadue,
Nucl. Phys. B \textbf{327}, 323 (1989).

\bibitem{Laenen:2000ij}
E.~Laenen, G.~F.~Sterman and W.~Vogelsang,
Phys. Rev. D \textbf{63}, 114018 (2001)
[arXiv:hep-ph/0010080 [hep-ph]].

\bibitem{Gehrmann:2010ue}
T.~Gehrmann, E.~W.~N.~Glover, T.~Huber, N.~Ikizlerli and C.~Studerus,
JHEP \textbf{06}, 094 (2010)
[arXiv:1004.3653 [hep-ph]].

\bibitem{Kramer:1996iq}
M.~Kramer, E.~Laenen and M.~Spira,
Nucl. Phys. B \textbf{511}, 523 (1998)
[arXiv:hep-ph/9611272 [hep-ph]].

\bibitem{Akhoury:1998tb}
R.~Akhoury, M.~G.~Sotiropoulos and G.~F.~Sterman,
[arXiv:hep-ph/9806388 [hep-ph]].

\bibitem{Kulesza:2002rh}
A.~Kulesza, G.~F.~Sterman and W.~Vogelsang,
Phys. Rev. D \textbf{66}, 014011 (2002)
[arXiv:hep-ph/0202251 [hep-ph]];
Phys. Rev. D \textbf{69}, 014012 (2004)
[arXiv:hep-ph/0309264 [hep-ph]].

\bibitem{Shimizu:2005fp}
H.~Shimizu, G.~F.~Sterman, W.~Vogelsang and H.~Yokoya,
Phys. Rev. D \textbf{71}, 114007 (2005)
[arXiv:hep-ph/0503270 [hep-ph]].

\bibitem{Laenen:2008ux}
E.~Laenen, L.~Magnea and G.~Stavenga,
Phys. Lett. B \textbf{669}, 173 (2008)
[arXiv:0807.4412 [hep-ph]].

\bibitem{Grunberg:2009yi}
G.~Grunberg and V.~Ravindran,
JHEP \textbf{10}, 055 (2009)
[arXiv:0902.2702 [hep-ph]].

\bibitem{Laenen:2010uz}
E.~Laenen, L.~Magnea, G.~Stavenga and C.~D.~White,
JHEP \textbf{01}, 141 (2011)
[arXiv:1010.1860 [hep-ph]].

\bibitem{Bonocore:2014wua}
D.~Bonocore, E.~Laenen, L.~Magnea, L.~Vernazza and C.~D.~White,
Phys. Lett. B \textbf{742}, 375 (2015)
[arXiv:1410.6406 [hep-ph]].

\bibitem{Bonocore:2015esa}
D.~Bonocore, E.~Laenen, L.~Magnea, S.~Melville, L.~Vernazza and C.~D.~White,
JHEP \textbf{06}, 008 (2015)
[arXiv:1503.05156 [hep-ph]].

\bibitem{Moch:2009mu}
S.~Moch and A.~Vogt,
JHEP \textbf{04}, 081 (2009)
[arXiv:0902.2342 [hep-ph]].

\bibitem{Vogt:2010cv}
A.~Vogt,
Phys. Lett. B \textbf{691}, 77 (2010)
[arXiv:1005.1606 [hep-ph]].

\bibitem{Almasy:2010wn}
A.~A.~Almasy, G.~Soar and A.~Vogt,
JHEP \textbf{03}, 030 (2011)
[arXiv:1012.3352 [hep-ph]].

\bibitem{LoPresti:2014ihe}
N.~A.~Lo Presti, A.~A.~Almasy and A.~Vogt,
Phys. Lett. B \textbf{737}, 120 (2014)
[arXiv:1407.1553 [hep-ph]].

\bibitem{Almasy:2015dyv}
A.~A.~Almasy, N.~A.~Lo Presti and A.~Vogt,
JHEP \textbf{01}, 028 (2016)
[arXiv:1511.08612 [hep-ph]].

\bibitem{DelDuca:2017twk}
V.~Del Duca, E.~Laenen, L.~Magnea, L.~Vernazza and C.~D.~White,
JHEP \textbf{11}, 057 (2017)
[arXiv:1706.04018 [hep-ph]].

\bibitem{Moult:2018jjd}
I.~Moult, I.~W.~Stewart, G.~Vita and H.~X.~Zhu,
JHEP \textbf{08} (2018), 013
[arXiv:1804.04665 [hep-ph]].

\bibitem{Ebert:2018gsn}
M.~A.~Ebert, I.~Moult, I.~W.~Stewart, F.~J.~Tackmann, G.~Vita and H.~X.~Zhu,
JHEP \textbf{04} (2019), 123
[arXiv:1812.08189 [hep-ph]].

\bibitem{Bahjat-Abbas:2019fqa}
N.~Bahjat-Abbas, D.~Bonocore, J.~Sinninghe Damst\'e, E.~Laenen, L.~Magnea, L.~Vernazza and C.~D.~White,
JHEP \textbf{11}, 002 (2019)
[arXiv:1905.13710 [hep-ph]].

\bibitem{vanBeekveld:2019cks}
M.~van Beekveld, W.~Beenakker, R.~Basu, E.~Laenen, A.~Misra and P.~Motylinski,
Phys. Rev. D \textbf{100}, no.5, 056009 (2019)
[arXiv:1905.11771 [hep-ph]].

\bibitem{vanBeekveld:2019prq}
M.~van Beekveld, W.~Beenakker, E.~Laenen and C.~D.~White,
JHEP \textbf{03}, 106 (2020)
[arXiv:1905.08741 [hep-ph]].

\bibitem{Ajjath:2020ulr}
A.~H.~Ajjath, P.~Mukherjee and V.~Ravindran,
[arXiv:2006.06726 [hep-ph]].

\bibitem{Ajjath:2020sjk}
A.~H.~Ajjath, P.~Mukherjee, V.~Ravindran, A.~Sankar and S.~Tiwari,
JHEP \textbf{04}, 131 (2021)
[arXiv:2007.12214 [hep-ph]].

\bibitem{Ajjath:2020lwb}
A.~H.~Ajjath, P.~Mukherjee, V.~Ravindran, A.~Sankar and S.~Tiwari,
Phys. Rev. D \textbf{103}, L111502 (2021)
[arXiv:2010.00079 [hep-ph]].

\bibitem{Ajjath:2021lvg}
A.~H.~Ajjath, P.~Mukherjee, V.~Ravindran, A.~Sankar and S.~Tiwari,
[arXiv:2107.09717 [hep-ph]].

\bibitem{Beneke:2019oqx}
M.~Beneke, A.~Broggio, S.~Jaskiewicz and L.~Vernazza,
JHEP \textbf{07}, 078 (2020)
[arXiv:1912.01585 [hep-ph]].

\bibitem{Broggio:2021fnr}
A.~Broggio, S.~Jaskiewicz and L.~Vernazza,
JHEP \textbf{10}, 061 (2021)
[arXiv:2107.07353 [hep-ph]].

\bibitem{Liu:2020tzd}
Z.~L.~Liu, B.~Mecaj, M.~Neubert and X.~Wang,
Phys. Rev. D \textbf{104} (2021) no.1, 014004
[arXiv:2009.04456 [hep-ph]].

\bibitem{Lustermans:2019cau}
G.~Lustermans, J.~K.~L.~Michel and F.~J.~Tackmann,
[arXiv:1908.00985 [hep-ph]].

\bibitem{Floratos:1981hs}
E.~G.~Floratos, C.~Kounnas and R.~Lacaze,
Nucl. Phys. B \textbf{192}, 417 (1981).

\bibitem{Mertig:1995ny}
R.~Mertig and W.~L.~van Neerven,
Z. Phys. C \textbf{70}, 637 (1996)
[arXiv:hep-ph/9506451 [hep-ph]].

\bibitem{Vogelsang:1995vh}
W.~Vogelsang,
Phys. Rev. D \textbf{54}, 2023 (1996)
[arXiv:hep-ph/9512218 [hep-ph]].

\bibitem{Vogelsang:1996im}
W.~Vogelsang,
Nucl. Phys. B \textbf{475}, 47 (1996)
[arXiv:hep-ph/9603366 [hep-ph]].

\bibitem{Moch:2014sna}
S.~Moch, J.~A.~M.~Vermaseren and A.~Vogt,
Nucl. Phys. B \textbf{889}, 351 (2014)
[arXiv:1409.5131 [hep-ph]].

\bibitem{Anastasiou:2003ds}
C.~Anastasiou, L.~J.~Dixon, K.~Melnikov and F.~Petriello,
Phys. Rev. D \textbf{69} (2004), 094008
[arXiv:hep-ph/0312266 [hep-ph]]. 
The analytic expression given in Eq.~(\ref{eqnnlo1}) may be extracted from 
the numerical code for this paper published at {\tt https://www.slac.stanford.edu/~lance/Vrap/}.

\bibitem{Anastasiou:2003yy}
C.~Anastasiou, L.~J.~Dixon, K.~Melnikov and F.~Petriello,
Phys. Rev. Lett. \textbf{91} (2003), 182002
[arXiv:hep-ph/0306192 [hep-ph]].

\bibitem{Hamberg:1990np}
R.~Hamberg, W.~L.~van Neerven and T.~Matsuura,
Nucl. Phys. B \textbf{359}, 343 (1991)
[erratum: Nucl. Phys. B \textbf{644}, 403 (2002)].

\bibitem{Ravindran:2006bu}
V.~Ravindran, J.~Smith and W.~L.~van Neerven,
Nucl. Phys. B \textbf{767} (2007), 100
[arXiv:hep-ph/0608308 [hep-ph]].

\bibitem{Catani:2009sm}
S.~Catani, L.~Cieri, G.~Ferrera, D.~de Florian and M.~Grazzini,
Phys. Rev. Lett. \textbf{103} (2009), 082001
[arXiv:0903.2120 [hep-ph]].

\bibitem{Gavin:2010az}
R.~Gavin, Y.~Li, F.~Petriello and S.~Quackenbush,
Comput. Phys. Commun. \textbf{182} (2011), 2388
[arXiv:1011.3540 [hep-ph]].

\bibitem{Chen:2021vtu}
X.~Chen, T.~Gehrmann, N.~Glover, A.~Huss, T.~Z.~Yang and H.~X.~Zhu,
Phys. Rev. Lett. \textbf{128}, no.5, 052001 (2022)
[arXiv:2107.09085 [hep-ph]].

\bibitem{prep} M.~Abele, D.~de~Florian, W.~Vogelsang, {\it work in preparation.}

\bibitem{Hou:2019efy}
T.~J.~Hou, J.~Gao, T.~J.~Hobbs, K.~Xie, S.~Dulat, M.~Guzzi, J.~Huston, P.~Nadolsky, J.~Pumplin and C.~Schmidt, \textit{et al.}
Phys. Rev. D \textbf{103}, no.1, 014013 (2021)
[arXiv:1912.10053 [hep-ph]].

\bibitem{Abdolmaleki:2021yjf}
H.~Abdolmaleki \textit{et al.} [xfitter Developers\textquoteright{} Team],
Phys. Rev. D \textbf{104}, no.5, 056019 (2021)
[arXiv:2105.11306 [hep-ph]].
 
\bibitem{Vogt:2000ci}
A.~Vogt,
Phys. Lett. B \textbf{497}, 228 (2001)
[arXiv:hep-ph/0010146 [hep-ph]].

\bibitem{KT} J.~Kodaira and L.~Trentadue, 
Phys.\ Lett.\ B {\bf 112}, 66 (1982); Phys.\ Lett.\ B {\bf 123}, 
335 (1983); S.~Catani, E.~D'Emilio and L.~Trentadue,
Phys.\ Lett.\ B {\bf 211}, 335 (1988).

\bibitem{Moch:2004pa} S.~Moch, J.~A.~M.~Vermaseren and A.~Vogt,
  Nucl.\ Phys.\ B {\bf 688}, 101 (2004)
  [hep-ph/0403192].

\bibitem{Harlander:2001is} R.~V.~Harlander and W.~B.~Kilgore,
Phys.\ Rev.\  D {\bf 64}, 013015 (2001)  [arXiv:hep-ph/0102241].

\bibitem{eric} T.~O.~Eynck, E.~Laenen and L.~Magnea,  
JHEP {\bf 0306}, 057 (2003)
[arXiv:hep-ph/0305179];
E.~Laenen and L.~Magnea,  
Phys.\ Lett.\  B {\bf 632}, 270 (2006) [arXiv:hep-ph/0508284].

\bibitem{Catani:2001ic}
S.~Catani, D.~de Florian and M.~Grazzini,
JHEP \textbf{05}, 025 (2001)
[arXiv:hep-ph/0102227 [hep-ph]].

\bibitem{Curci:1980uw}
G.~Curci, W.~Furmanski and R.~Petronzio,
Nucl. Phys. B \textbf{175}, 27 (1980).
 

\end{thebibliography}
\end{document}